\documentclass[apj]{emulateapj}
\usepackage{apjfonts}

\def\chandra    {\emph{Chandra}}
\def\xmm        {\emph{XMM}}
\def\vla        {\emph{VLA}}
\def\rosat        {\emph{ROSAT}}
\def\gmrt {\emph{GMRT}}

\def \deg      {$^{\circ}$}

\def\arcsec{$^{\prime\prime}$}

\def\lax{\lesssim}

\def\pcmsq{cm$^{-2}$}

\shorttitle{Recurrent radio outbursts in NGC\,1407}
\shortauthors{S.~Giacintucci et al.}

\begin{document}

\setlength{\pdfpageheight}{\paperheight}
\setlength{\pdfpagewidth}{\paperwidth}

\slugcomment{{\em The Astrophysical Journal}, accepted on 2012 June 23}

\title{Recurrent radio outbursts at the center of the NGC\,1407 galaxy
  group}

\author{Simona Giacintucci\altaffilmark{1,2}, 
Ewan O'Sullivan\altaffilmark{3},
Tracy E. Clarke\altaffilmark{4},
Matteo Murgia\altaffilmark{5},
Jan M. Vrtilek\altaffilmark{3},
Tiziana Venturi\altaffilmark{6},
Laurence P. David\altaffilmark{3},
Somak Raychaudhury\altaffilmark{7},
Ramana M. Athreya\altaffilmark{8}
}
\altaffiltext{1}{Department of Astronomy, University of Maryland,
  College Park, MD 20742, USA; simona@astro.umd.edu}
\altaffiltext{2}{Joint Space-Science Institute, University of Maryland, College Park,
MD, 20742-2421, USA}
\altaffiltext{3}{Harvard-Smithsonian Center for Astrophysics, 
60 Garden Street, Cambridge, MA 02138, USA}
\altaffiltext{4}{Naval Research Laboratory, Code 7213, Washington, DC
  20375, USA}
\altaffiltext{5}{INAF-Osservatorio Astronomico di Cagliari,
  Loc. Poggio dei Pini, Strada 54, I-09012 Capoterra (CA), Italy}
\altaffiltext{6}{INAF - Istituto di Radioastronomia, via Gobetti 101, I-40129 Bologna,
Italy}
\altaffiltext{7}{School of Physics and Astronomy, University of Birmingham, Birmingham,
B15 2TT, UK}
\altaffiltext{8}{Indian Institute of Science Education and Research,
  Central Tower, Sai Trinity Building, Sutarwadi Road, Pashan, Pune 411021, India}

\begin{abstract}
We present deep {\em Giant Metrewave Radio Telescope} (\gmrt)
radio observations at 240, 330 and 610 MHz of the complex 
radio source at the center of the NGC1407 galaxy group. Previous \gmrt\ 
observations at 240 MHz revealed faint, diffuse
emission enclosing the central twin-jet radio galaxy. This has been
interpreted as an indication of two possible radio outbursts occurring at 
different times. Both the inner double and diffuse component are detected 
in the new \gmrt\ images at high levels of significance. Combining
the \gmrt\ observations with archival {\em Very Large
  Array} data at 1.4 and 4.9 GHz, we derive the total spectrum of
both components. The inner double has a spectral index
$\alpha=0.7$, typical for active, extended radio galaxies, whereas
the spectrum of the large-scale emission is very steep, with
$\alpha=1.8$ between 240 MHz and 1.4 GHz. The radiative 
age of the large-scale component is very long, $\sim 300$ Myr,
compared to $\sim$30 Myr estimated for the central double, confirming
that the diffuse component was generated during a former cycle
of activity of the central galaxy. The current activity 
have so far released an energy which is nearly one order of magnitude 
lower than that associated with the former outburst. The group 
X-ray emission in the \chandra\ and \xmm-Newton images and extended 
radio emission show a similar swept-back morphology. We speculate 
that the two structures are both affected by the motion of the group 
core, perhaps due to the core sloshing in response to a
recent encounter with the nearby elliptical galaxy NGC\,1400.
\end{abstract}


%%%%%%%%%%%%%%%%%%%%%%%%%%%%%%%%%%%%%%%%%%%%%%%%%%
% SECTION 1: Introduction
%%%%%%%%%%%%%%%%%%%%%%%%%%%%%%%%%%%%%%%%%%%%%%%%%%%%

\section{Introduction}
\label{sec:intro}

Quoting \cite{1999A&A...348..699L}, {\em ``radio sources are born, grow and
finally ... sleep''}.  The dormant phase in the evolutionary 
course of a radio galaxy and fate of the radio-emitting plasma
after the cessation of the nuclear activity are among the most 
intriguing issues in extragalactic astronomy. 
The active stage of a powerful radio source, typically associated 
with an elliptical galaxy, can last several $10^7$ up to few $10^8$
years. During this time, the radio galaxy is likely fed by mass
accretion onto the supermassive black hole of the host galaxy. 
Once this accretion is interrupted, or becomes insufficient to support 
radio activity, the radio source enters a dying phase
\citep[e.g.,][and references therein]{2011A&A...526A.148M}; 
the radio emission passively evolves and rapidly fades, even if
expansion losses are negligible and the relativistic electrons are
subject only to radiative losses. This is reflected into a pronounced steepening
of the integrated radio spectrum whose slope $\alpha$ can reach
ultra-steep values, $\alpha \gtrsim 2$ \citep[e.g.,][]{1994A&A...285...27K},
adopting the convention $S_{\nu} \propto \nu^{-\alpha}$ for the
synchrotron spectrum, where $S_{\nu}$ is the flux density at the frequency $\nu$.
Eventually, the radio emission disappears below the detection limit of
present radio telescopes. 
 
During the fading stage of a radio source, 
the central nucleus may switch on again and produce 
new radio activity, thus leading to a restarted radio source. 
Evidence for episodic radio activities are reported in the 
literature for a growing number of radio galaxies (see, for
instance, Saikia \& Jamrozy 2009 for a review).
Double-double radio galaxies \citep{2000MNRAS.315..371S}
qualify as one of the most unmistakable examples of recurrent 
nuclear activity; here a new pair of inner lobes are produced 
close to the nucleus before the previously generated, more 
distant ones have completely faded
\citep[e.g.,][]{1996MNRAS.279..257S, 1999A&A...348..699L, 2009BASI...37...63S}.
Another type of restarted radio sources are
``nesting'' radio galaxies \citep{2007MNRAS.378..581J},
i.e., sources characterized by a faint, extended 
structure, within which a bright 
radio source, extended on a significantly smaller scale, is embedded.
Examples are Hercules A
\citep{2003MNRAS.342..399G,2005MNRAS.358.1061G},
3C\,310 \citep{1984ApJ...282L..55V,1986MNRAS.222..753L},
and 4C\,29.30 \citep{2007MNRAS.378..581J}.
In few well-studied cases, the large--scale emission and 
inner source appear clearly, and quite sharply, separated in the spectral 
index distribution \citep[e.g.,][]{2003MNRAS.342..399G},
with much steeper spectral index values measured 
for the outer emission. This has been interpreted as 
evidence for different epochs of jet activity, with
the more extended and steeper component being
related to an earlier outburst of the central active galactic nucleus (AGN).

In this paper, we focus on the complex radio source associated 
with the elliptical galaxy NGC\,1407 (see Table \ref{tab:ngc1407}), 
at the center of the poor group of galaxies Eridanus~A at z=0.0059,
with a one-dimensional velocity dispersion $\sigma_{v} =$372 km~s$^{-1}$ 
\citep{2006MNRAS.369.1351B} and 
0.3-2 keV X-ray luminosity of $10^{41.7}$ erg~s$^{-1}$
\citep{2006PASA...23...38F}.
Based on {\em Giant 
Metrewave Radio Telescope} (\gmrt) observations at 240 MHz  
and 610 MHz, Giacintucci et al. (2011, hereinafter G11)
%\cite{2011ApJ...732...95G}(hereinafter G11)
suggested that this source may be a restarted radio galaxy. Here, we present a 
multifrequency radio study of NGC\,1407 based on new, deep \gmrt\ 
observations at 240 MHz, 330 MHz and 610 MHz, and multifrequency 
data from the {\em Very  Large Array} (\vla) archive. Our study
is complemented with the analysis of {\em Chandra} and {\em
  XMM-Newton} X-ray data.

\section{Radio observations} \label{sec:radioobs}

In this section, we present the new \gmrt\ observations of 
NGC\,1407 at 240 MHz, 330 MHz and 610 MHz. Details 
on these observations are summarized in Table \ref{tab:gmrt} 
which reports: observing date, frequency and total bandwidth
in columns 1, 2 and 3; total time on source (column 4); 
full-width half maximum (FWHM) and position angle (PA) of the full 
array (column 5); rms level (1$\sigma$) at full resolution (column 6). 

The \gmrt\ data were supplemented by all useful observations in
the \vla\ archive. The observing details are provided in 
Table \ref{tab:vla}, which shows the project code and array
configuration in the first two columns; the other columns provide the
same information as Table \ref{tab:gmrt}.

%
%%%%%%%%%%%% Tab. 1 - General properties %%%%%%%%%%%%%%%%%%%%%%
%
\begin{table}
\caption[]{General properties of NGC\,1407}
\begin{center}
%\footnotesize
\begin{tabular}{lc}
\hline\noalign{\smallskip}
\hline\noalign{\smallskip}
RA$_{\rm J2000}$ (h m s) & 03 40 11.9 \\
DEC$_{\rm J2000}$ ($^{\circ}$ $^{\prime}$ 
                     $^{\prime\prime}$) & $-$18 34 39 \\
z & 0.0059 \\
$D_{L}$ (Mpc) & 25.0 \\
angular scale (kpc/ \arcsec) & 0.120 \\
morphology & E0 \\
$M_R$ (mag) & $-$22.8  \\
\noalign{\smallskip}
\hline\noalign{\smallskip}
\end{tabular}
\end{center}
Notes to Table \ref{tab:ngc1407} -- RA$_{\rm J2000}$ and DEC$_{\rm J2000}$ 
are the coordinates of the optical galaxy from the
NASA/IPAC Extragalactic Database (NED). The optical redshift $z$ is
from \cite{1994A&A...283..722Q}. The galaxy morphological
classification is from the Third Reference 
Catalogue of Bright Galaxies \citep{1991rc3..book.....D}.
The Cousins R-band absolute magnitude is from
\cite{2006MNRAS.369.1375T}. We assume a flat cosmology with 
H$_0$ = 71 km s$^{-1}$ Mpc$^{-1}$, $\Omega_{\lambda}$ = 0.73 and
$\Omega_{0}$ = 0.27.
\label{tab:ngc1407}
\end{table}
%
%%%%%%%%%%%% end of Tab. 2  %%%%%%%%%%%%%%%%%%%%%%%%%
%

%%%%%%%%%%%% Tab. 2: - GMRT observations %%%%%%%%%%%%%%%%%%%%%%%%%
%
\begin{table*}[htbp!]
\caption[]{Details of the \gmrt\ observations.}
\begin{center}
\footnotesize
\begin{tabular}{cccccccc}
\hline\noalign{\smallskip}
\hline\noalign{\smallskip}
Observation & $\nu$ & $\Delta \nu$  & t  & FWHM,
p.a. & rms \\ %& {\it u-v} range & LDS  \\
date & (MHz)& (MHz) & (min) & ($^{\prime
     \prime}\times^{\prime \prime}$, $^{\circ}$)&  (mJy
   beam$^{-1}$) \\
\noalign{\smallskip}
\hline\noalign{\smallskip}
Nov 19, 2009 & \phantom{0}240 $^a$& 8 & 270 & 16.1$\times$10.9, 36 & 0.25  \\
 Jan 17, 2010 & 330 & 32& 330 &  12.1$\times$9.2, 42   & 0.16   \\
Nov 19, 2009 & \phantom{0}610 $^a$& 32 & 270  & 8.2$\times$4.4, 46 & 0.05\\
\noalign{\smallskip}
\hline\noalign{\smallskip}
\end{tabular}
\end{center}
Notes to Table 2 -- $a$: observed in simultaneous 240 MHz/610
MHz mode.
\label{tab:gmrt}
\end{table*}
%
%%%%%%%%%%%% end of Tab. 2  %%%%%%%%%%%%%%%%%%%%%%%%%
%

%
%%%%%%%%%%%% Tab. 3 - VLA observations %%%%%%%%%%%%%%%%%%%%%%%%%
%
\begin{table*}[htbp!]
\caption[]{Details of the \vla\ archive observations.}
\begin{center}
\footnotesize
\begin{tabular}{cccccccccc}
\hline\noalign{\smallskip}
\hline\noalign{\smallskip}
Project & Array & Observation & $\nu$ & \phantom{0}$\Delta \nu$  & t  & FWHM,
p.a. & rms \\ %& {\it u-v} range & LDS  \\
code & & date & (MHz)& (MHz) & (min) & ($^{\prime
     \prime}\times^{\prime \prime}$, $^{\circ}$)&  (mJy
   beam$^{-1}$) \\
\noalign{\smallskip}
\hline\noalign{\smallskip}
AS827  & B  & Apr 30, 2005 & 322/329 & 6.3 &  130 & 27.7$\times$17.0,
$-$16 & 3 \\
AV151 &  A & July 3, 1987 & 1446/1496 & 12.5 & 3 & 2.2$\times$1.3,
11 & 0.1 \\
AR536 & CD & May 28, 2004 & 1385/1465 & \phantom{0}50 & 20 &
76.2$\times$20.8, 54 & 0.08 \\
AW136 & BC& Jun 23, 1985 & 4835/4885 & \phantom{0}50 & 6.5 & 5.1$\times$2.0, 71 & 0.14\\
AW112 &  C  & Jun 2, 1984   & 4835/4885 & \phantom{0}50 & 8.5 & 15.9$\times$4.4, 44 & 0.07\\
\noalign{\smallskip}
\hline\noalign{\smallskip}
\end{tabular}
\end{center}
\label{tab:vla}
\end{table*}
%
%%%%%%%%%%%% end of Tab. 3 %%%%%%%%%%%%%%%%%%%%%%%%%
%

\subsection{\gmrt\/ observations}
\label{sec:gmrtobs}

NGC\,1407 was observed with the \gmrt\ at 240 MHz, 330 MHz 
and 610 MHz (project 17$_{-}$034; Table \ref{tab:gmrt}) 
in a full-synthesis run of approximately 9 hours (including 
calibration overheads) at each frequency. The observations at 
240 MHz and 610 MHz were made using the dual-frequency mode, 
recording LL polarization at 240 MHz and RR polarization at 610 MHz. 
Both RR and LL were recorded at 330 MHz. 

The 330 MHz and 610 MHz data were collected using the upper 
and lower side bands simultaneously (USB and LSB, respectively), 
for a total observing bandwidth of 32 MHz. Only the USB was used 
at 240 MHz, with 8 MHz bandwidth. The default spectral-line observing 
mode was used, with 128 channels for each band (64 channel at 240 MHz) 
and a spectral resolution of 125 kHz/channel. The datasets were
calibrated and reduced using the NRAO Astronomical Image Processing 
System package (AIPS). We refer to 
Giacintucci et al.\ (2008, 2011) for a description of the data reduction.

Self-calibration was applied to reduce residual phase variations and
improve the quality of the final images. Due to the large field of
view of the \gmrt\ at all frequencies, we used the wide-field imaging 
technique at each step of the phase self-calibration process, to
account for the non-planar nature of the sky. The final images were 
produced using the multi-scale CLEAN implemented 
in the AIPS task IMAGR, which results in better imaging of extended sources 
compared to the traditional CLEAN \citep[e.g.,][for a detailed 
discussion, see Appendix A in \cite{2009AJ....137.4718G}]{2006AJ....131.2900C}.
We used delta functions as model components for the unresolved
features and circular Gaussians for the resolved ones, with increasing
width to progressively highlight the extended emission during the clean.
Beyond the image at full resolution (Table \ref{tab:gmrt}), we produced
images with lower resolution (down to $\sim 45''$), tapering the $uv$ data
by means of the parameters ROBUST and UVTAPER in IMAGR. The final
images were corrected for the \gmrt\ primary 
beam response using the task PBCOR in AIPS. 
The rms noise level (1$\sigma$) achieved in the final 
images at full resolution is given in Table \ref{tab:gmrt}.

We adopted the scale of \cite{1977A&A....61...99B}
for the flux density calibration. Residual amplitude errors are within
$8\%$ at 240 MHz and 330 MHz, and $5\%$ at 
610 MHz \citep{2004ApJ...612..974C}.

\subsection{\vla\/ archive data}
\label{sec:vlaobs}

We extracted and reduced \vla\ archive data of NGC\,1407 at 327 MHz, 
1.4 GHz and 4.9 GHz (see Table \ref{tab:vla} for details). The
observations were all pointed on NGC\,1407, excepting those at 
4.9 GHz in BnC configuration, where the phase center is 6$^{\prime}$ 
south-west of the galaxy.

The observations at 327 MHz were obtained in multi-channel
continuum mode with 16 channels and a total bandwidth of 6.3 MHz. 
The dataset was calibrated using 3C48 as bandpass and amplitude
calibrator. Due to problems with the source observed as phase
calibrator (0340-213), 3C48 was also used for the phase
calibration. 

All data were processed in AIPS and images were produced using
the standard Fourier transform deconvolution method.
Self-calibration was applied to reduce the effects of residual
phase errors in the data. Due to the large field of view at 327 MHz, 
the wide-field imaging technique was implemented to correct
for distortions in the image caused by the non-coplanarity of the
\vla. Correction for the primary beam attenuation was applied 
to the final images at all frequencies using the task PBCOR in AIPS. 
All flux densities are on the \cite{1977A&A....61...99B} 
scale and average residual amplitude errors are $\lax 5$\%.

\section{The radio images}\label{sec:images}

The \gmrt\ 330 MHz image at full resolution 
($\sim$12$^{\prime\prime}$) is presented in Fig.~\ref{fig:hr}a,
overlaid on the optical image from The Second Palomar Sky 
Survey (POSS 2). Faint, diffuse emission (white contours) is detected 
around the central double source S3 (magenta contours),
in good agreement with the 240 MHz image in G11 (also shown
in Fig.~\ref{fig:xmm}). S1, S2 and S4 are discrete radio sources 
located, in projection, within the diffuse emission. S1 and S2 
have no optical counterpart on the POSS--2 image, while
a very faint object seems to be associated with S4, suggesting
that all three sources are background radio galaxies.

Hereinafter, we will refer to the two components of 
the NGC\,1407 system as the inner double and large-scale emission.

\subsection{The inner double}

\begin{figure*}[ht!]
\epsscale{1.2}
\plotone{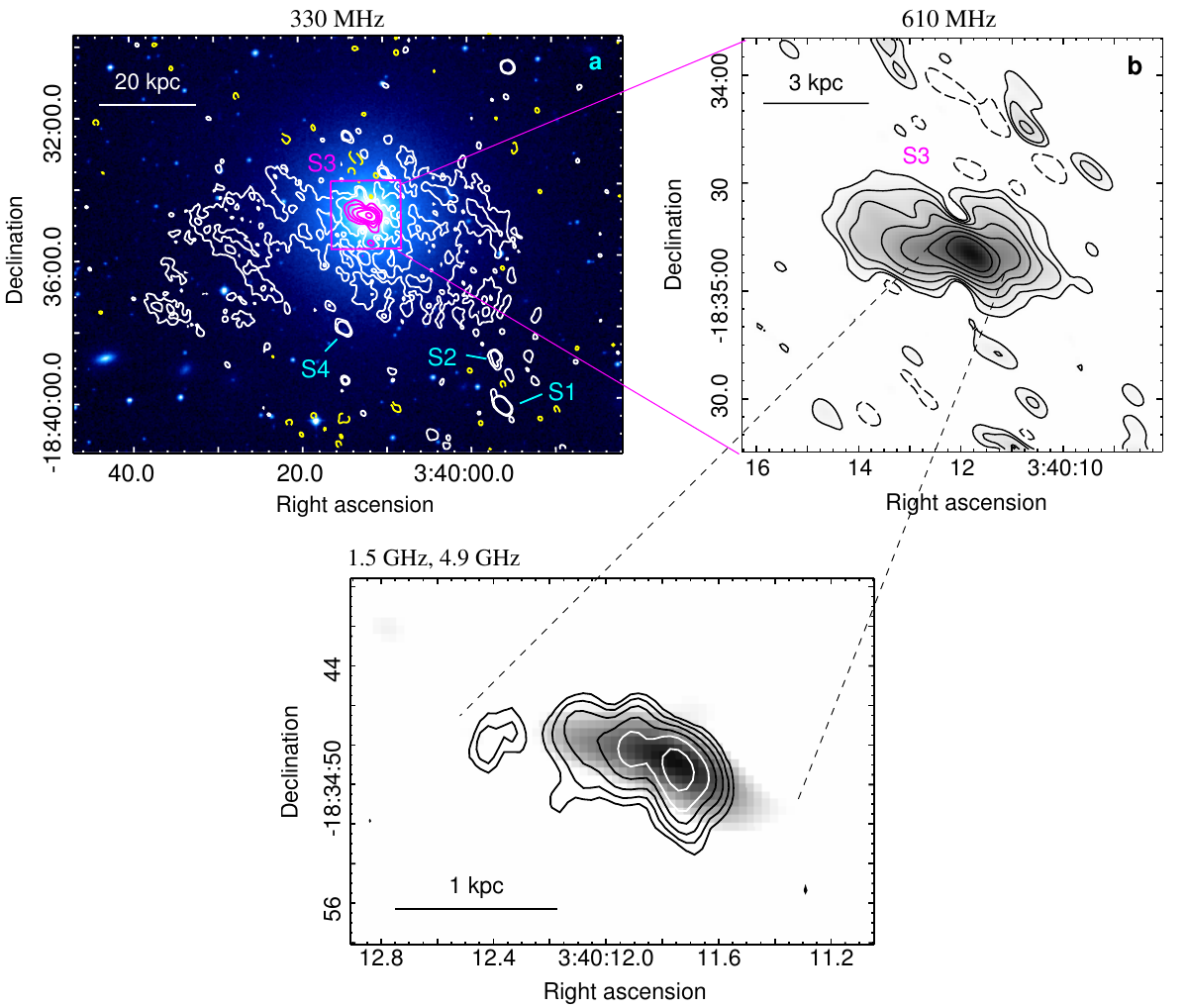}
\caption{{\em a}: \gmrt\ 330 MHz full resolution contours (white,
  magenta and yellow), overlaid on the optical POSS--2 image. The restoring 
beam is $12.1^{\prime \prime}\times9.2^{\prime\prime}$, p.a. 42$^{\circ}$. The rms
noise level is 0.2 mJy beam$^{-1}$. Contour levels are -0.6 (dashed
yellow), 0.6, 1.2, 2.4 (white), 4.8, 9.6, 20 and 80 (magenta) mJy beam$^{-1}$. 
Labels indicate the discrete radio sources embedded in the extended
structure. {\em  b}: \gmrt\ 610 MHz full resolution image (contours and gray
scale) of the inner double (S3).
The restoring beam is $8.2^{\prime \prime}\times4.4^{\prime\prime}$,
p.a. 46$^{\circ}$. The rms noise level is 0.05 mJy beam$^{-1}$. 
Contour levels are spaced by a factor of 2 starting from 0.2 mJy
beam$^{-1}$. Dashed contours correspond to $-0.2$ mJy beam$^{-1}$.
 {\rm c:} \vla\ 1.5 GHz contours (FWHM=
2.2$^{\prime\prime}\times1.3^{\prime\prime}$, $1\sigma$=0.1 mJy beam$^{-1}$) 
overlaid on the \vla\  gray-scale image at 4.9 GHz (FWHM= 
5.1$^{\prime\prime}\times2.0^{\prime\prime}$, $1\sigma$=0.14 mJy 
beam$^{-1}$). Contours scale by a factor of 2 starting from 0.3 
mJy beam$^{-1}$.}
\label{fig:hr}
\end{figure*}

Fig.~\ref{fig:hr}b zooms on the central double source (S3). The region
corresponds to the magenta box in Fig.~\ref{fig:hr}a.
The contours and gray-scale image are the emission detected at 610 
MHz at full resolution ($8^{\prime\prime}\times4^{\prime\prime}$).
Thanks to the higher sensitivity, more extended emission associated
with the double is visible in this new 610 MHz image, compared to the
previous observations published in G11 (also reported in Fig.~\ref{fig:chandra}).

The 2$^{\prime\prime}$-resolution image
from the \vla--A data at 1.5 GHz 
(Fig.~\ref{fig:hr}c, contours) shows the
innermost structure of the source. The 
contours are overlaid on the \vla--BnC
image at 4.9 GHz at resolution
5$^{\prime\prime}\times2^{\prime\prime}$.
At $\sim$1 kpc scale, the source appears a 
core-jet with position angle of $\sim$30$^{\circ}$, which is 
approximately aligned with the axis of the double on larger scale
(panel b). A faint blob of 
emission is visible to the east at 1.5 GHz, 
apparently detached from the core-jet. 
The blob is not detected at 4.9 GHz.

\subsection{The large-scale emission}

\begin{figure*}[ht!]
\epsscale{0.8}
\plotone{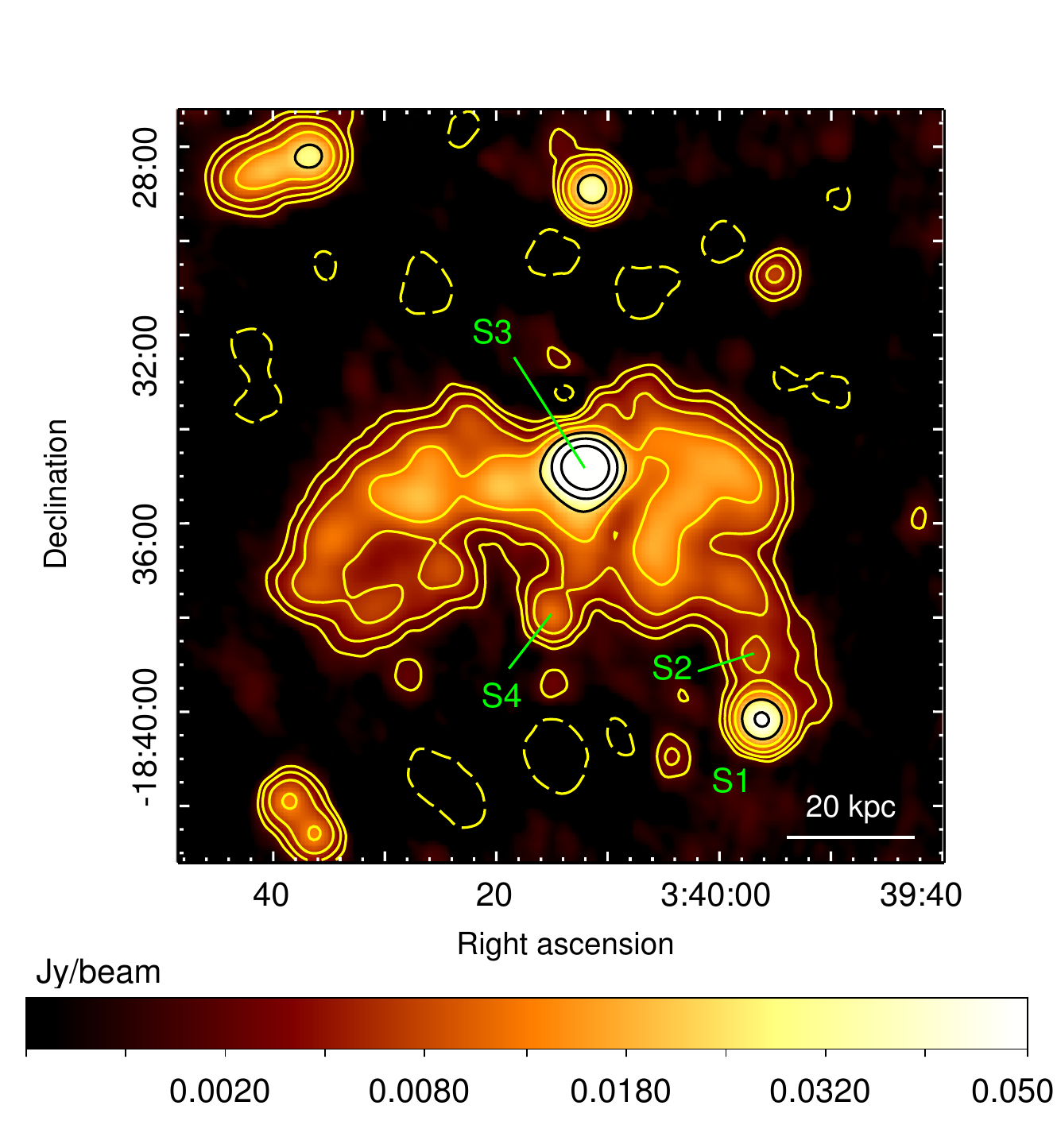}
\caption{\gmrt\ low-resolution image at 330 MHz. The image has
been restored with a beam of $45.0^{\prime
  \prime}\times45.0^{\prime\prime}$, p.a. 0$^{\circ}$. The rms noise
level is 0.5 mJy beam$^{-1}$. Contours start at +3$\sigma$ and 
then scale by a factor of 2. The $-3\sigma$ level is shown by the
dashed contours. Labels indicate the discrete 
radio sources (Fig.~\ref{fig:hr}{\em a}).}
\label{fig:327lr}
\end{figure*}

The \gmrt\ 330 MHz image, produced at the lower resolution 
of 45$^{\prime\prime}\times$45$^{\prime\prime}$, is presented in 
Fig.~\ref{fig:327lr}. The new images at 240 MHz 
and 610 MHz are shown at similar resolution in Fig.~3 and Fig.~4a.
The large-scale emission is detected at high levels 
of significance at all frequencies. Its total extent is $\sim$80 kpc, 
consistent in all images.
The brightest regions of this component are also visible in the 
\vla--CnD image at 1.4 GHz, restored with a circular beam 
of 60$^{\prime\prime}$ (Fig.~4b).
It is clear that the large-scale emission becomes progressively 
fainter with increasing frequency. 
The morphology also seems to change from the 1.4 GHz
image, where the source shows a bipolar structure with east-west axis,
to the 240 MHz image, where the extended emission becomes
fairly amorphous and more extended in the north-south direction. 
This suggests that the radio spectrum of this component
is very steep.

%
%%%%%%%%%%%%%%%%%%%%%% Fig. 1 %%%%%%%%%%%%%%%%%%%%%%%%%
%
%
%%%%%%%%%%%%%%%%%% end of Fig. 1 %%%%%%%%%%%%%%%%%%%%%%%%%
%

%
%%%%%%%%%%%%%%%%%%%%%% Fig. 2 %%%%%%%%%%%%%%%%%%%%%%%%%
%
%
%%%%%%%%%%%%%%%%%%%%%% End of Fig. 2 %%%%%%%%%%%%%%%%%%%%%%%%%
%

%
%%%%%%%%%%%%%%%%%%%%%% Fig. 3 %%%%%%%%%%%%%%%%%%%%%%%%%
%
\begin{figure*}[ht!]
\epsscale{0.8}
\plotone{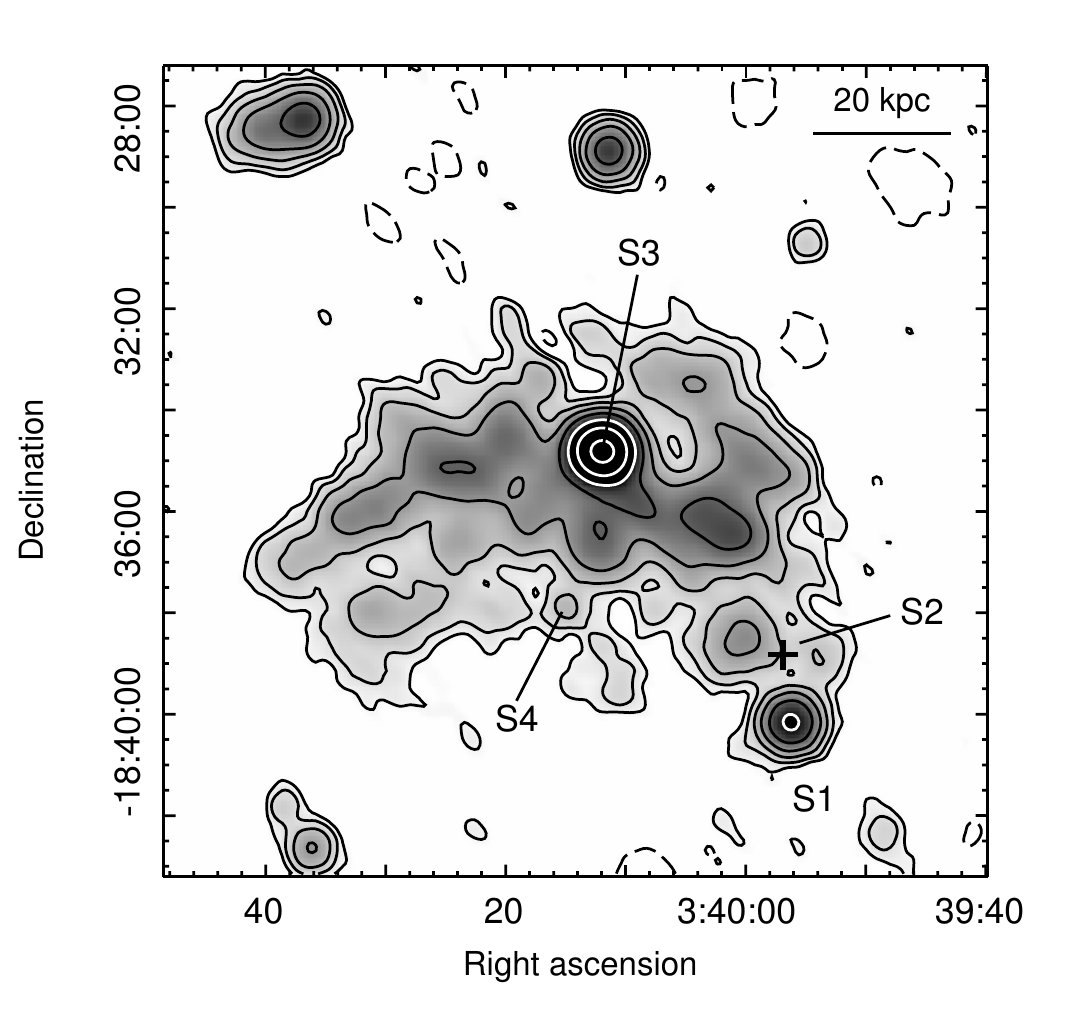}
\caption{\gmrt\ low-resolution image (contours and gray scale)
at 240 MHz. The restoring beam is $45.0^{\prime
  \prime}\times45.0.^{\prime\prime}$, p.a. 0$^{\circ}$. 
The rms noise level is 1 mJy beam$^{-1}$.
Contours start at +3$\sigma$ and then scale by 
a factor of 2. The $-3\sigma$ level is shown by the
dashed contours.  Labels indicate the discrete 
radio sources (Fig.~\ref{fig:hr}{\em a}).}
\label{fig:235lr}
\end{figure*}
%
%%%%%%%%%%%%%%%%%%%%%% End of Fig. 3 %%%%%%%%%%%%%%%%%%%%%%%%%
%

%
%%%%%%%%%%%%%%%%%%%%%% Fig. 4 %%%%%%%%%%%%%%%%%%%%%%%%%
%
\begin{figure*}[ht!]
\epsscale{1.15}
\plottwo{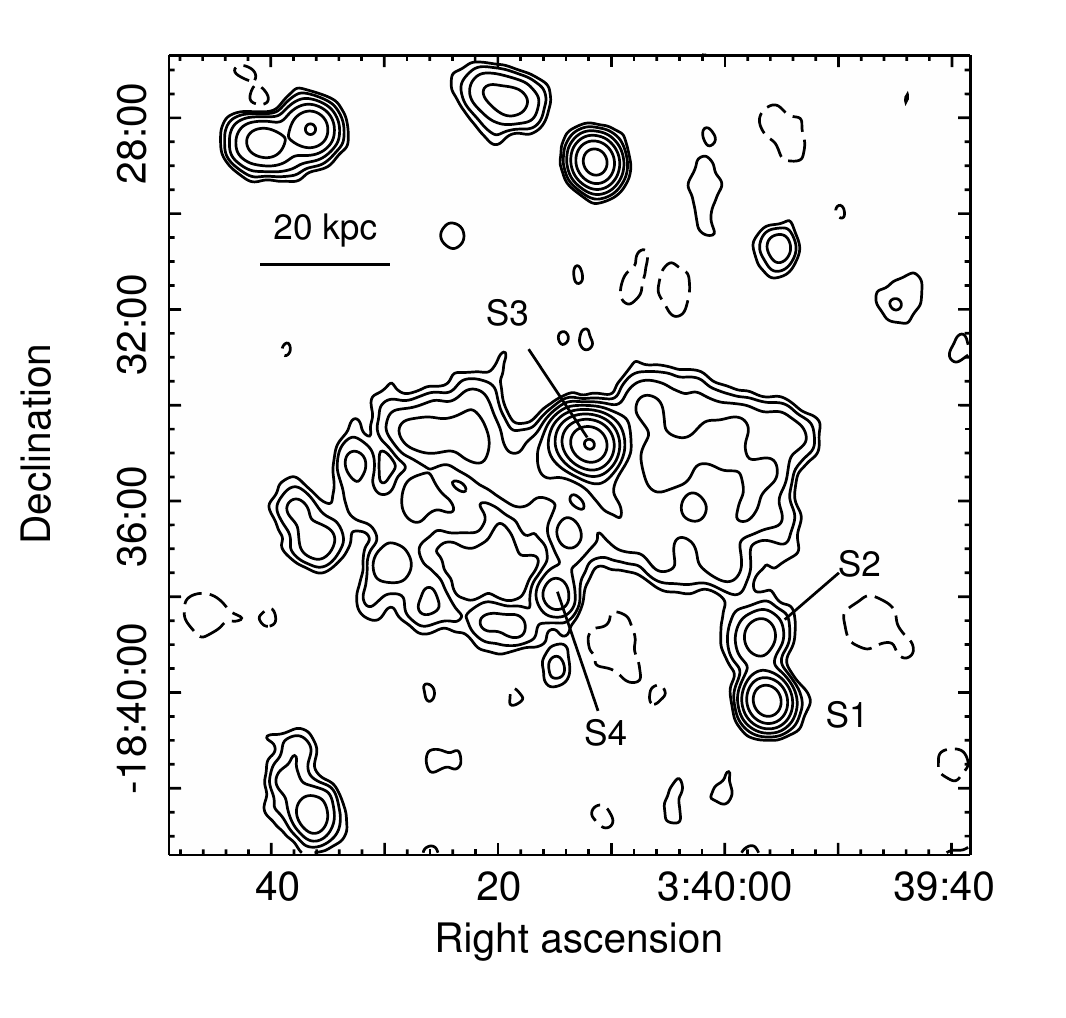}{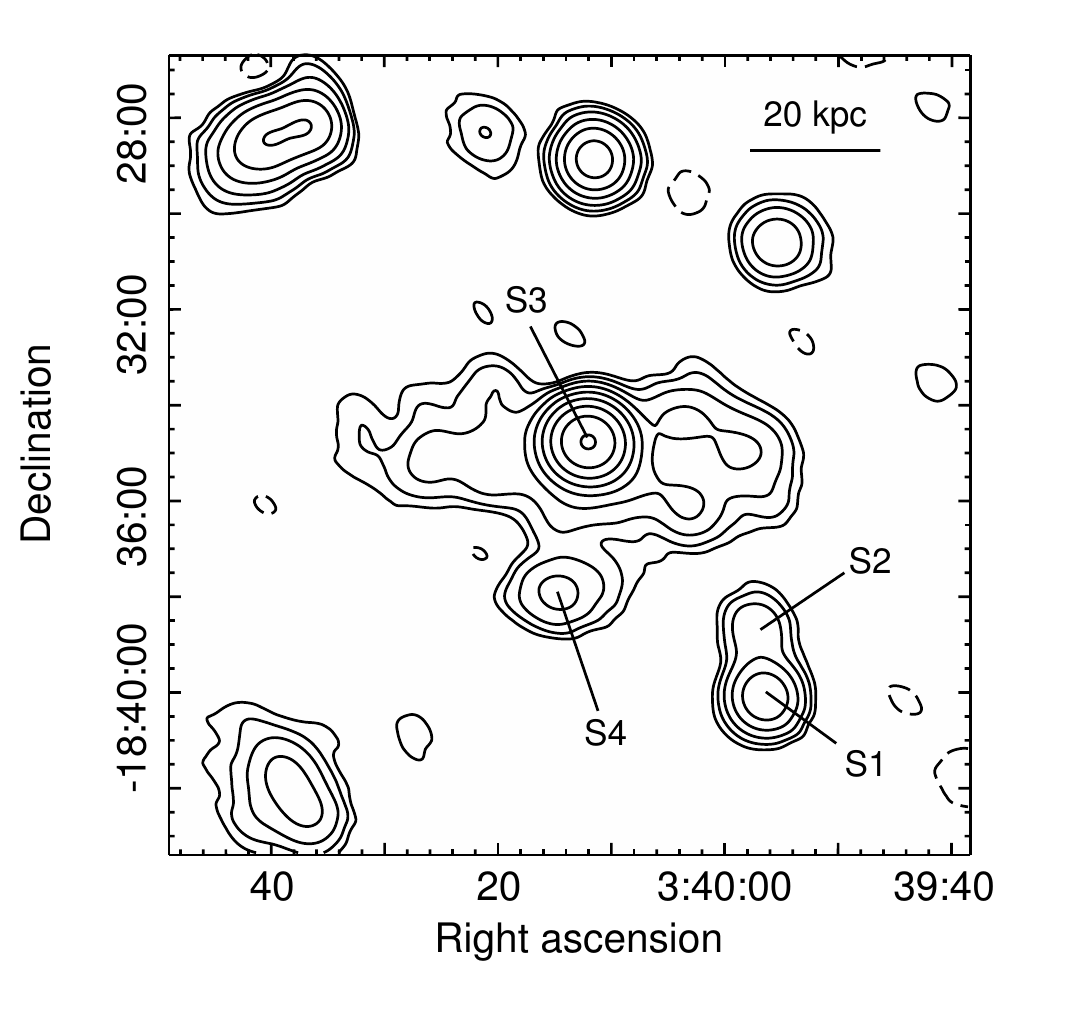}
\caption{{\em a:} \gmrt\ low resolution contours at 610 MHz. 
The restoring beam is $41.8^{\prime
  \prime}\times37.0.^{\prime\prime}$, p.a. 17. 
The rms noise level is 0.15 mJy beam$^{-1}$.
{\em b:} {\em VLA--CnD} low resolution contours at 1425 MHz.
The image has been restored with a circular beam of $60^{\prime
  \prime}.$  The rms noise level is 0.08 mJy beam$^{-1}$.
In both panels contours start at +3$\sigma$ and then scale by 
a factor of 2. The $-3\sigma$ level is shown as dashed contours.
Labels indicate the discrete radio sources (Fig.~\ref{fig:hr}{\em a}).}
\label{fig:610lr}
\end{figure*}
%
%%%%%%%%%%%%%%%%%%%%%% end of Fig. 4 %%%%%%%%%%%%%%%%%%%%%%%%%
%

\section{Radio spectral properties, magnetic fields and radiative age}
\label{sec:spectra}

We studied the spectral properties -- total integrated spectrum and
spectral index distribution -- and derived the physical parameters
of NGC\,1407 using the Synage++ package (Murgia 2001).
 
\begin{itemize}
\item[1.] We fitted the radio spectrum of the inner double and
large-scale emission separately, and derived the injection spectral index
 $\alpha_{\rm inj}$ and break frequency $\nu_{\rm break}$ for each
  component.

\item[2.] We obtained a spectral index image of the system
and extracted the distribution of $\alpha$ along the source structure
as a function of the distance from the center. By fitting the observed
spectral index trend, we derived  $\nu_{\rm break}$ for the
large-scale emission.

\item[3.] We calculated the magnetic field $B_{\rm eq}$ 
assuming that the relativistic particle and magnetic field energy 
densities are in approximate energy equipartition. We assumed cylindrical 
geometry, a filling factor of unity and that there is equal energy in
relativistic ions and electrons ($k=1$). We also assumed that the
magnetic field is unordered along the line of sight. We used the 
radio luminosity at 240 MHz, $\alpha_{\rm  inj}$ from the spectral
modelling, and imposed a low-energy cutoff 
of $\gamma_{\rm min}$=100 in the energy distribution 
of the radiative electrons, where $\gamma$ is the electron Lorentz
factor \citep[e.g.,][]{1997A&A...325..898B,2005AN....326..414B}.
This allows us to take into account the contribution from relativistic
electrons with an energy as low as $\sim$50 MeV.

\item[4.] Under a number of assumptions, the knowledge of the break 
frequency in the spectrum of a radio source allows us to estimate the time 
elapsed since the source formation \citep[e.g.,][]{1985ApJ...291...52M}.
We assumed that radiative (i.e., synchrotron and inverse 
Compton) losses dominate over expansion losses, and that the magnetic
field strength is uniform across the source and remains constant over the source
life time. We also assume that the relativistic electron population is 
isotropic \citep{1973A&A....26..423J}
and reacceleration processes can be neglected. We used 
the following equation

$$ t_{\rm rad} = 1590  \frac{B_{\rm eq}^{0.5}}{(B_{\rm eq}^2 + B_{\rm CMB}^2)} [(1+z) \nu_{\rm break}]^{-0.5} $$

\noindent to derive the radiative age $t_{\rm rad}$
of both radio components,
where $t_{\rm rad}$ is expressed in Myr,
$\nu_{\rm break}$ in GHz, and $B_{\rm eq}$ and $B_{\rm
  CMB}$ in $\mu$G. $B_{\rm CMB} = 3.2(1+z)^2$ is the equivalent
magnetic field of the cosmic microwave background (CMB) radiation,
i.e., the magnetic field strength with energy density equal to that of
the CMB at the redshift $z$.

\end{itemize}

Our analysis is presented in detail in the following subsections, and 
results are summarized in Table \ref{tab:phypar}, which provides the radio luminosity 
at 240 MHz, volume, observed spectral index, radiative model used for
the spectral fitting, injection spectral index, break 
frequency, equipartition magnetic field, radiative age and
equipartition pressure for each component.

\subsection{Integrated spectra}\label{sec:sp}

We measured the flux density of the inner double and
large--scale emission in similar regions at all frequencies, 
i.e., the extent of the double at 610 MHz (Fig.~1b) and area 
covered by the diffuse emission at 240 MHz (Fig.~3). 
The flux densities are summarized in Table \ref{tab:fluxes}, 
along with the associated uncertainties (1$\sigma$) and 
angular resolution of the images used for the measurements.

The \vla\ image at 327 MHz (not shown here) detects the 
inner double, but does not show significant diffuse emission around it,
due to its lower sensitivity compared to the \gmrt\ 330 MHz image (Tab.~3). 
For this reason, Table \ref{tab:fluxes} provides only the flux density
at 327 MHz for the central double. 

We also inspected the VLSS\footnote{\vla\ Low-frequency 
Sky Survey \citep{2007AJ....134.1245C}} image at 74 MHz and found 
a weak source at the position of the inner double, with a peak
flux density of $\sim$400 mJy beam$^{-1}$. The source is not listed in
the VLSS catalog, implying that its total flux density at 74
MHz is less than the $5\sigma$ local rms level (1$\sigma$=86 mJy
beam$^{-1}$). We therefore assume an upper limit of $S_{\rm 74 \,
MHz}$=430 mJy for the double. No evidence of significant emission over the
large--scale structure is visible on the VLSS image. Given that the 
total area occupied by the diffuse component corresponds 
to $\sim$24 VLSS beams (each beam is 80$^{\prime \prime}\times
80^{\prime \prime}$), we can place a $3\sigma$ upper limit of 6 Jy at 74 MHz. 

In Fig.~\ref{fig:spectra} we show the integrated radio spectra of the 
inner double (red points) and large-scale emission (black points). 
The data points at 1.5 GHz and 4.9 GHz from A and BnC configurations
(Tables 2 and \ref{tab:fluxes}) are not shown in the plot; due to the
high angular resolution and lack of short spacings, the images from 
these datasets detect only the innermost region of the double (Fig.~1c). 

The overall spectrum of the inner double has $\alpha_{\rm obs}=0.69\pm0.03$ between 
240 MHz and 4.9 GHz. A spectral steepening seems to occur at high
frequency, with $\alpha_{\rm obs}$ changing from $0.65\pm0.05$ below 1425 MHz to
$0.75\pm0.06$ above. The large--scale emission has a much
steeper spectrum, with $\alpha_{\rm obs}=1.80\pm0.05$ in the 240 MHz-1.4 GHz range. 
The upper limit at 74 MHz suggests a low-frequency
flattening of the spectrum, with $\alpha_{\rm obs} \lax 1.6$ below 240 MHz.

\subsubsection{Spectral modelling}\label{sec:spmod}

We fitted the spectrum of the inner double using 
a continuous injection (CI) model \citep[e.g.,][]{1962SvA.....6..317K}, in which the
source is continuously replenished by a constant flow of fresh relativistic
electrons with a power-law energy distribution in a region of constant magnetic
field. Under these assumptions, the radio spectrum has a standard
shape, with a low-frequency spectral index representing the
$\alpha_{\rm inj}$  of the youngest electron population, and a
high-frequency spectral index limited to be
$\le \alpha_{\rm inj} +0.5$ at frequencies above $\nu_{\rm break}$.

\begin{figure}[ht!]
%\epsscale{0.7}
\epsscale{1.2}
\plotone{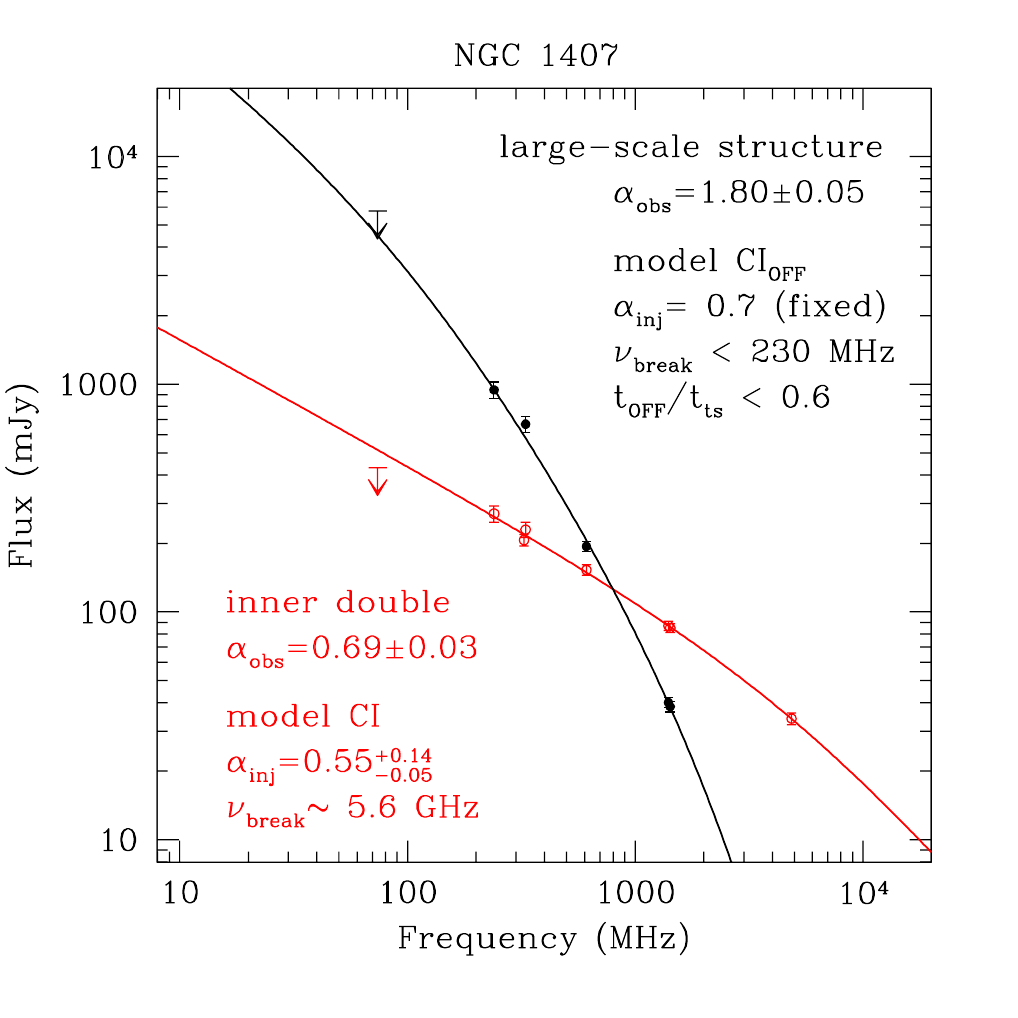}
\caption{
Integrated spectra of the inner double (red circles) and diffuse
component (black circles) using the flux density reported in 
Table \ref{tab:fluxes}. The solid lines are the best-fit CI (red) and
CI$_{\rm OFF}$ (black) models described in the text (Sect. \ref{sec:spmod}). }
\label{fig:spectra}
\end{figure}

The resulting best-fit for the inner double (red line in
Fig.~\ref{fig:spectra}) yields a non-aged spectral index $\alpha_{\rm
  inj}$=0.55$^{+0.14}_{-0.05}$ and $\nu_{\rm  break} \sim 5.6$ GHz. 
With these values, the estimated equipartition magnetic field and
radiative age of the central double are $B_{\rm eq} \sim$7 $\mu$G  and
$t_{\rm rad} \sim 30$ Myr (Table 5).

The very steep spectral index of the large-scale component ($\alpha_{\rm
  obs}=1.8$) suggests that this emission is old. For this reason, we
modelled its spectrum using a CI$_{\rm OFF}$ model
\citep[e.g.,][]{2007A&A...470..875P,2011A&A...526A.148M}. This model 
assumes an initial phase of electron
injection at a constant rate by the nuclear source (CI phase), followed by
a switch-off of the nuclear activity. A dying phase then begins, with
the radio emission rapidly fading subject to the energy losses of the
relativistic electrons. It is assumed that the magnetic field
strength is uniform within the source, energy losses are dominant
with respect to other processes, and the pitch angle of the radiating
electrons has an isotropic distribution with respect to the local direction
of the magnetic field.  We imposed an initial spectral index $\alpha_{\rm
  inj}$=0.7, similar to the observed $\alpha$ of the inner double (Tab. 5),
and derived an upper limit of 230 MHz for the break frequency.  The
estimated magnetic field is $B_{\rm eq} \sim$2.1 $\mu$G and the total
radiative lifetime is very long, i.e., $t_{\rm rad} >320$ Myr.

The CI$_{\rm  OFF}$ model also provides $t_{\rm OFF}/t_{\rm rad}$,
i.e., the dying to total source age ratio \citep[for details see][]{2007A&A...470..875P}. 
We found that the active phase lasted for at least 
$128$ Myr and the nuclear activity switched off not more than 
$t_{\rm off} \sim 192$ Myr ago.

%
%%%%%%%%%%%%%%%%%%%%%%%% Table 4 - flux densities %%%%%%%%%%%%%%%%
%
\begin{table}
\caption[]{Flux densities of the NGC\,1407 system}
\begin{center}
\begin{tabular}{lccl}
\hline\noalign{\smallskip}
\hline\noalign{\smallskip}
component  & $\nu$   &  $S_{\nu}$ & FWHM \\
                   & (MHz)    &  (mJy)  & (\arcsec $\times$ \arcsec) \\
\hline\noalign{\smallskip}
inner double (S3)\dotfill  
& 74  &  $< 430$   &  80$\times$80 (VLSS) \\
& 240  & 270$\pm$22 & 15.4$\times$12.5 (\gmrt\, this work) \\
& 327  & 207$\pm$12 & 28$\times$17 (\vla\, this work) \\ 
& 330  & 230$\pm$18& 12.0$\times$9.2 (Fig.~\ref{fig:hr}{\em a}) \\
& 610  & 153$\pm$8&   8.2$\times$4.4 (Fig.~\ref{fig:hr}{\em b})\\
& 1400 & 87$\pm$4 & 45$\times$45 (NVSS) \\
&1425 & 85$\pm$4 & 60$\times$60 (Fig.~4b) \\
& 1.50  & 37$\pm$2& 2.2$\times$1.3  (Fig.~\ref{fig:hr}{\em c}) \\
& 4860  & 34$\pm$2& 15.9$\times$4.4 (\vla\, this work)\\
& 4860  & 23$\pm$1& 5.1$\times$2.0  (Fig.~\ref{fig:hr}{\em c}) \\
&&&\\
large-scale 
emission & 74 & $< 6000$ & 80$\times$80 (VLSS) \\
& 240 & 945$\pm$76 & 45.0$\times$45.0 (Fig.~\ref{fig:235lr}) \\
 & 330 & 668$\pm$53   & 45.0$\times$45.0  (Fig.~\ref{fig:327lr}) \\
& 610 & 194$\pm$10   & 41.8$\times$37.0  (Fig.~\ref{fig:610lr}{\em a})\\
& 1425 & 38$\pm$2& 60.0$\times$60.0  (Fig.~\ref{fig:610lr}{\em b})\\
 \hline\noalign{\smallskip}
\end{tabular}
\end{center}
Notes -- The flux density of the large-scale emission has
been obtained after subtraction of the contribution of the inner
double (S3) and point sources S1, S2 and S4.
\label{tab:fluxes}
\end{table}
%
%%%%%%%%%%%%%%%%%%%%%%End of Tab. 4 %%%%%%%%%%%%%%%%%%%%%%%%%%%%%%%%%%%%%

%
%%%%%%%%%%%%%%% FIG 5 - radio spectrum %%%%%%%%%%%%%%%%%%%%%%%%
%

%
%%%%%%%%%%%%%%% End of Fig. 5 %%%%%%%%%%%%%%%%%%%%%%%%%%%%%%%%%
%

%
%%%%%%%%%%%%%%% Tab 5- fit results %%%%%%%%%%%%%%%%%%%%%%%%
%

%{\rotate
\begin{table*}[htbp]
%\footnotesize
\arraycolsep=0mm
\caption[]{Spectral modelling results, equipartition magnetic fields and radiative ages}
\begin{center}
\begin{tabular}{lccccccccccccc}
\hline\noalign{\smallskip}
\hline\noalign{\smallskip}
 & $L_{\rm 240 \, MHz}$ & V & $\alpha_{\rm obs}$ & spectral & $\alpha_{\rm inj}$ &
 $\nu_{\rm break}$ & $B_{\rm eq}$ & %$u_{\rm min}$ & 
$t_{\rm rad}$ & $t_{\rm OFF}/t_{\rm rad}$ & $t_{\rm CI}$ & $t_{\rm
  OFF}$ & $P_{\rm radio}$\\
&  (W Hz$^{-1}$) &  (kpc$^3$) & & model & & (GHz) & ($\mu$G) & %(erg cm$^{-3}$) & 
(Myr) & & (Myr) & (Myr)  & (erg cm$^{-3}$) \\
\hline\noalign{\smallskip}

inner double & $2.02\times10^{22}$& 140 &  $0.69\pm0.03$ & CI & 0.55$^{+0.14}_{-0.05}$
&  \phantom{00}5.6 &  7 &  \phantom{00}30 &  $-$ & $-$ & $-$ & 2.8$\times$10$^{-12}$\\
&&&&&&&&&&&&\\
large-scale & $7.07\times10^{22}$ & $84\times10^3$ & $1.80\pm0.05$ & CI$_{\rm
  OFF}$ $^a$ &  0.7 $^c$& $<0.23$ & 2.1 & $> 320$ & $< 0.60$ &
$>128$ & $<192$ &  2.5$\times$10$^{-13}$\\
emission & & & & CI$_{\rm
  OFF}$ $^b$ &  0.7 $^c$ & \phantom{00}0.34 & &  \phantom{00}264
&   \phantom{00}0.65 &  \phantom{00}92 &  \phantom{00}172 & \\
 \hline\noalign{\bigskip}
\end{tabular}
\end{center}
\label{tab:phypar}
Notes to Table 5-- 
$a$: fit of the total integrated spectrum; $b$: fit of the spectral index trend; $c$: fixed.   
\end{table*}
%}
%
%%%%%%%%%%%%%%%%%%%%%%End of Tab. 5%%%%%%%%%%%%%%%%%%%%%%%%%%%%%%%%%%%%%

\subsection{Spectral index image}\label{sec:spix}

\begin{figure*}
\epsscale{1.1}
\plottwo{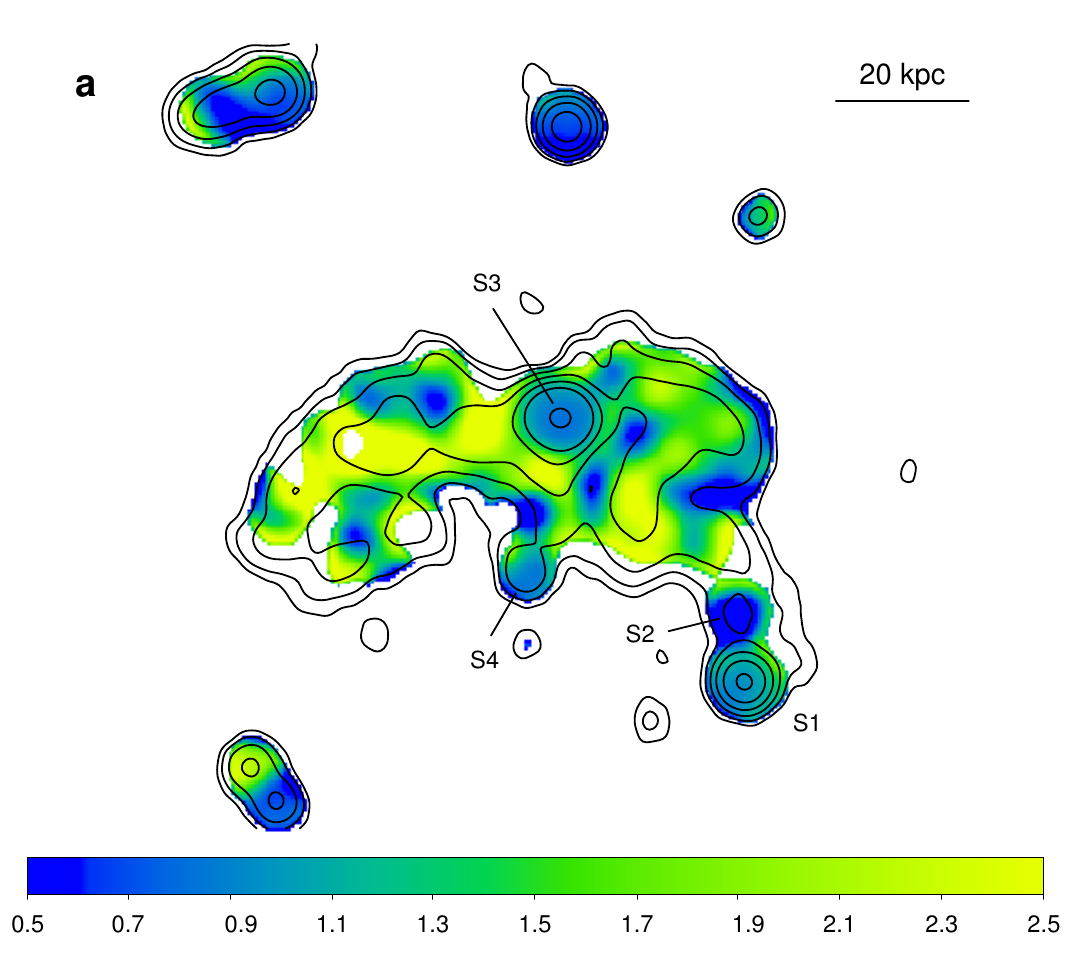}{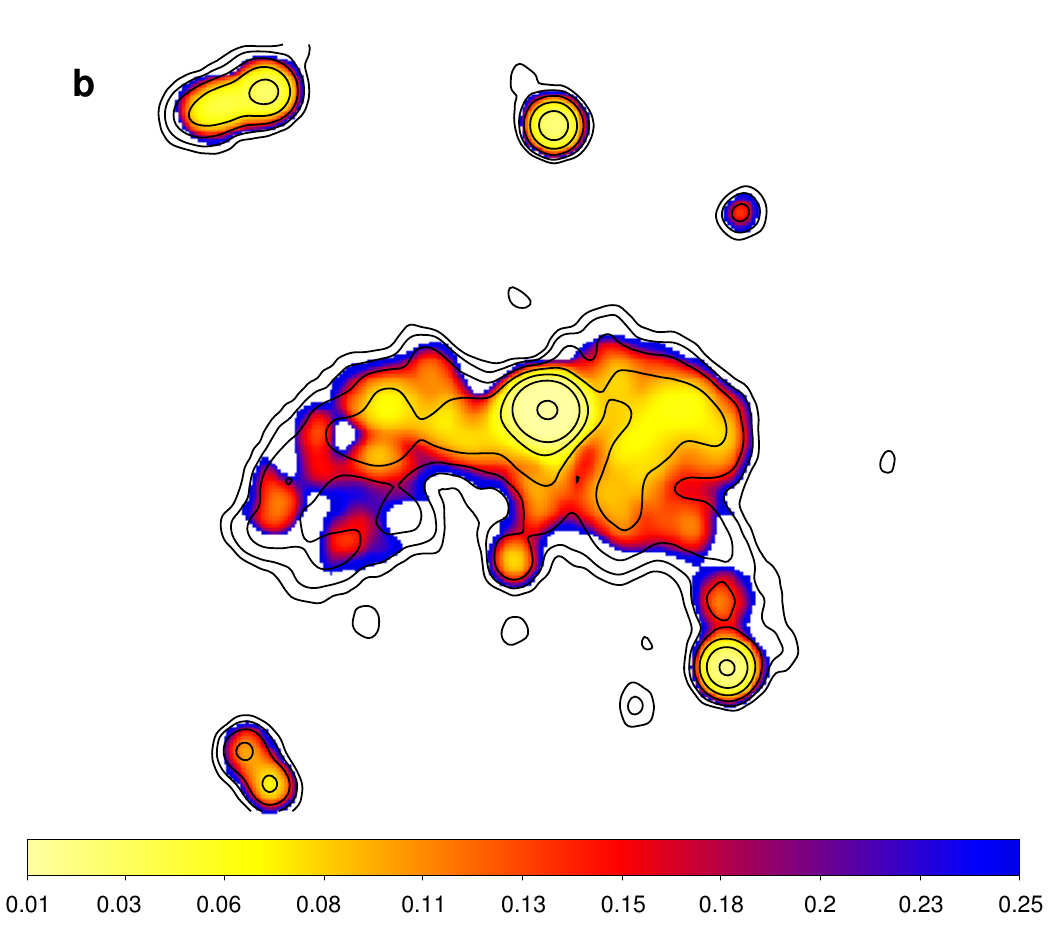}
\caption{Color scale image of the spectral index (a) and spectral
index uncertainty (b) distributions between 330 MHz and 610 MHz. The
image has been computed from images with a restoring beam of $45^{\prime
  \prime}\times45^{\prime \prime}$. Contours at 330 MHz and labels are as in 
Fig.~2. The spectral index uncertainty map is based on the rms noise
in both radio images. The absolute flux calibration uncertainty is not included.
}
\label{fig:spix}
\end{figure*}

We obtained images of the radio spectral index distribution
by comparing sets of \gmrt\ images at different 
frequencies, produced with the same cell size, $u–v$ range, 
restoring beam, and corrected for the primary beam attenuation. The images were
aligned, the pixels with brightness below the 5$\sigma$ level were blanked,
and finally the images were combined into a spectral index
image. 
In Fig.~\ref{fig:spix}a we present our best spectral index image 
obtained by comparing the images at 330 MHz and 610 MHz, both 
restored with a circular beam of $45^{\prime  \prime}$ radius and with 
noise levels similar to the images in Fig.~2 and Fig.~4a. We also 
show the distribution of the spectral index uncertainty based on 
the rms noise in both radio images (Fig.~\ref{fig:spix}b) and overlay 
the 330 MHz contours on both images to provide a reference for the 
source morphology. We note that Fig.~\ref{fig:spix}b does not include the 
absolute flux scale uncertainty of 5\% at 610 MHz and 8\% at 330 MHz, 
which results in a typical error on the spectral index of $\sigma_{\alpha} = 0.15$.

%
%%%%%%%%%%%%%%% FIG 6 - spectral index %%%%%%%%%%%%%%%%%%%
%
%
%%%%%%%%%%%%%%%%% End of Fig. 6 %%%%%%%%%%%%%%%%%%%%%%%
%

The inner double (S3), unresolved at this resolution, has an average
spectral index of $\sim 0.75\pm0.02$, consistent with
$\alpha_{\rm obs}= 0.66\pm0.15$ given by the flux densities measured 
at the two frequencies (Table \ref{tab:fluxes}). Sources S1 and S4 have both 
$\alpha \sim 0.8$, while the weak source S2 has
 $\alpha \sim$0.6. The large-scale component has spectral
index values ranging between $\alpha\sim 1.5$ and $\alpha\sim 2.5$,
with uncertainties of $\sim 0.06-0.08$ in the brightest regions 
and up to $\sim$0.25 at the edges. 

\begin{figure}
\epsscale{1.2}
%\epsscale{0.8}
\plotone{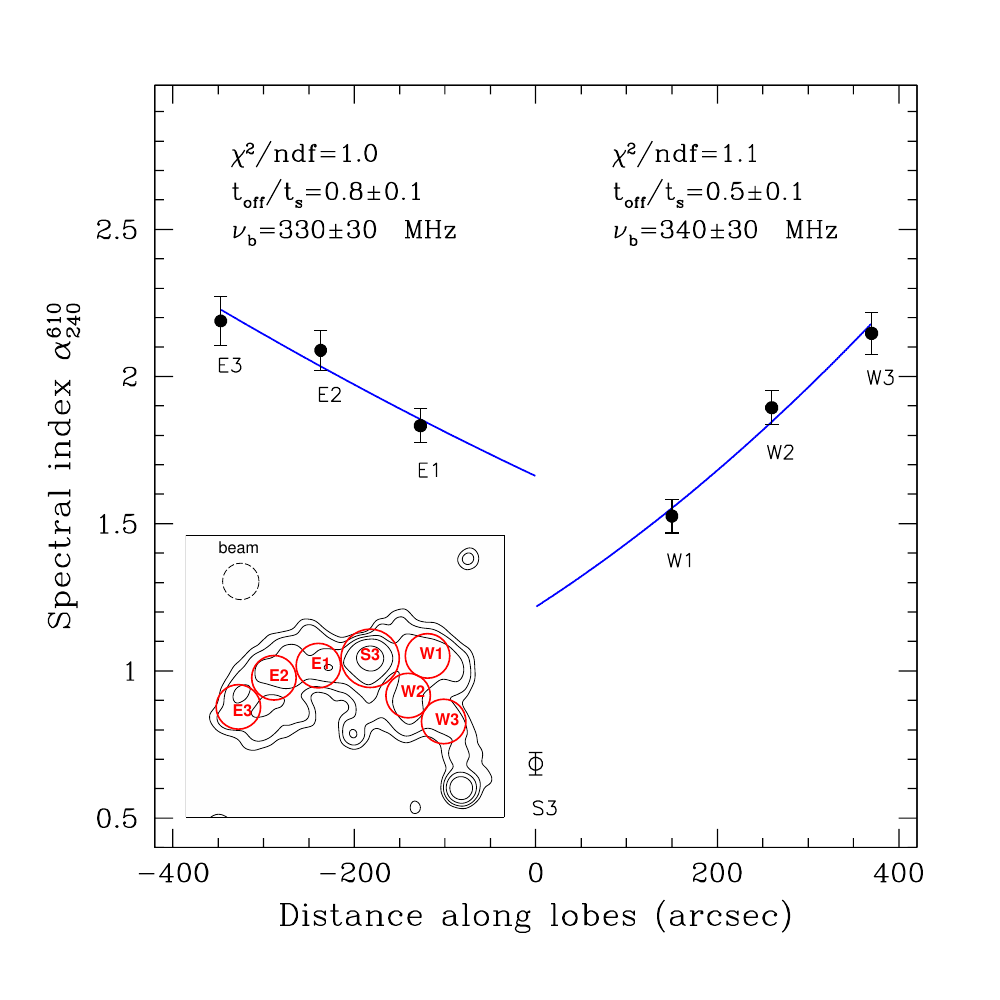}
\caption{240 MHz-610 MHz spectral index distribution as function of
  the distance from the center, computed using the  55$^{\prime
  \prime}$-radius circular boxes shown in the inset. The solid line
is the best fit of the radiative model described in the text. }
\label{fig:steep}
\end{figure}

Fig.~\ref{fig:steep} shows the spectral index profiles along the
eastern and western portions of the diffuse emission. We first 
measured the flux density in circular regions on the individual images
at 240 MHz and 610 MHz (see inset), both restored with a 
circular beam of $45^{\prime\prime}$-radius, and then computed the 
corresponding spectral index. The circles were placed following
the brightest regions along the diffuse emission. The radius of each
circle was set to 55$^{\prime  \prime}$ to ensure a sufficient
signal-to-noise ratio and independent data points. In the plot we 
also show the data point corresponding to the inner double
S3, obtained from the same images. As noticed  in Section \ref{sec:sp}, 
there is a clear and abrupt separation in $\alpha$ between S3 and
outer emission, the latter being substantially steeper, with $
\alpha=1.5\pm0.1$ to $2.2\pm0.1$
along the western region and $\alpha=1.8\pm0.1$ to
$2.2\pm0.1$ in the eastern one.

%
%%%%%%%%%%%%%%  FIG 7 - spectral index trend %%%%%%%%%%%%%%%%%%%
%
%
%%%%%%%%%%%%%%%% End of Fig. 7 %%%%%%%%%%%%%%%%%%%%%%%%%%%%
%

We interpreted the spectral index trend in Fig.~\ref{fig:steep} in terms of
radiative aging of the relativistic electrons by synchrotron and
inverse Compton processes \citep[e.g.,][and references
therein]{2007A&A...470..875P}, assuming that radiative losses dominate 
over expansion losses and reacceleration. If the magnetic field and
expansion velocity of the relativistic
plasma are both constant, we can then estimate the 
minimum break frequency from the fit of the observed spectral
index profiles. We used the relation $\nu_{\rm break} \propto
d^{-2}$, where $d$ is the distance from the core, which 
reflects the fact that the radiating electrons age as they travel away from 
the nucleus.
The two profiles along the large-scale component 
were fitted separately adopting a CI$_{\rm OFF}$ model (Sect. \ref{sec:spmod})
with $\alpha_{\rm inj}$ fixed to 0.7, i.e., the spectral index of S3.
The best fit, shown as blue lines in Fig.~\ref{fig:steep}, gives a
similar break frequency in the east and west portions of the source --
 $\nu_{\rm break} = 330\pm30$ MHz and $\nu_{\rm break} = 340\pm30$
 MHz, respectively -- and a ratio $t_{\rm off}/t_{\rm rad}$ of 0.6 and
 0.5. Using the average between these values and a magnetic field
 strength of 2.1 $\mu$G (Table 5), we estimated a total age 
of $\sim 264$ Myr. The CI phase lasted approximately 92 Myr and 
the nuclear engine switched off $\sim$172 Myr ago.

The total age derived here is shorter than the lower limit of
$\sim$320 Myr obtained from the fit of the total 
spectrum (Table 5). A reason for such discrepancy may lie in the
use of different areas to extract the flux densities in the two
cases; the region used to derive the total spectrum corresponds 
to the whole area occupied by the diffuse emission at 240 MHz, 
including the emission which is the most distant from the center and, 
thus, plausibly the oldest. The circular regions in
Fig.~\ref{fig:steep} sample instead only a part of that region. 

Another consideration is that the spectral index between 240 MHz and 610 MHz 
($\alpha_{\rm obs}=1.70\pm0.10$) is slightly flatter than the average
slope of the total spectrum,
i.e., $\alpha_{\rm obs}=1.80\pm0.05$ between 240 MHz and 1.4 GHz. This could
result in a break frequency from the spectral trend fitting
which is higher than that estimated from the total spectrum, thus yielding
a shorter age (for the same magnetic field strength).

Finally, it is important to bear in mind that both methods rely on
the age determination from the break frequency, which has large
uncertainties due to the number of underlying assumptions (see
discussion in Section \ref{sec:age}).

%
%%%%%%%%%%%%% FIG 8 - XMM %%%%%%%%%%%%%%%%%%%%%%%%%%%%%%%%%
%
\begin{figure*}[ht!]
\epsscale{1}
\plotone{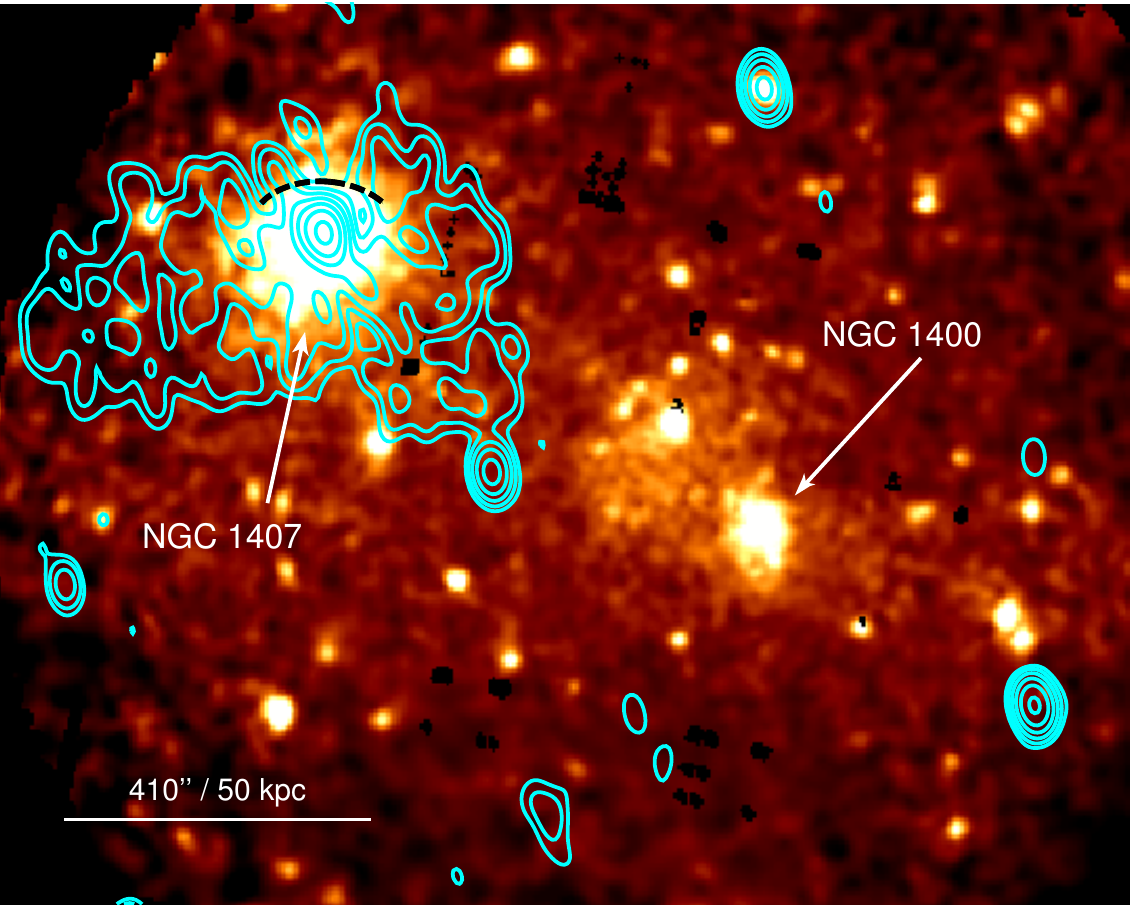}
\caption{{\em XMM-Newton} adaptively smoothed 0.3-2 keV image of the
  NGC\,1407 group, with \gmrt\ 240 MHz contours overlaid from G11
  (FWHM=$48.5^{\prime\prime}\times31.9^{\prime\prime}$). Contours
  start at $3\sigma=3$ mJy beam$^{-1}$, and then scale by a factor of
  2. NGC\,1400, a group member, is marked, as well as the trail of
  X-ray emission associated with it. The black dashed arc marks the 
surface brightness edge in Fig.~\ref{fig:chandra}.}
\label{fig:xmm}
\end{figure*}
%
%%%%%%%%%%%%%%%% End of Fig. 8 %%%%%%%%%%%%%%%%%%%%%%%%%%%%%
%

%
%%%%%%%%%%%%% FIG 9 - Chandra %%%%%%%%%%%%%%%%%%%%%%%%%%%%%%%%%
%
\begin{figure*}[ht!]
\epsscale{1}
\plottwo{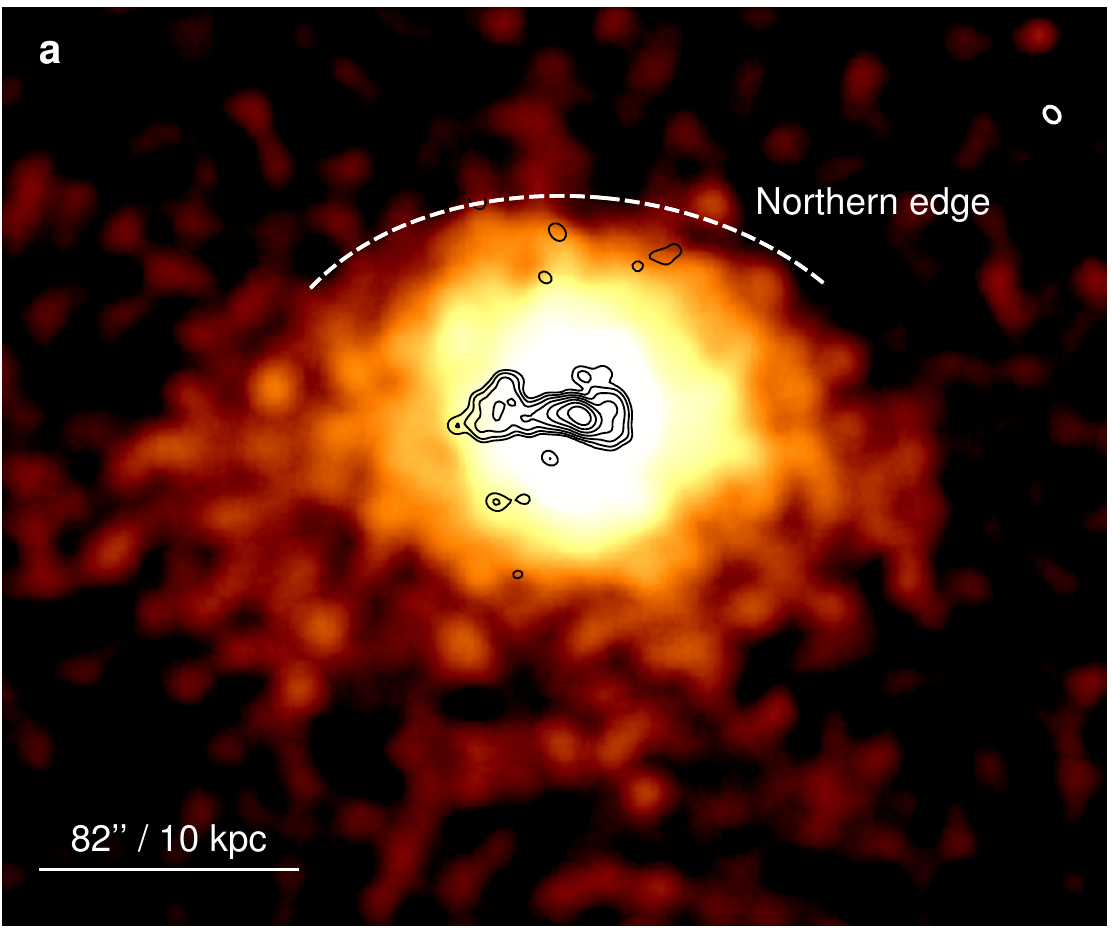}{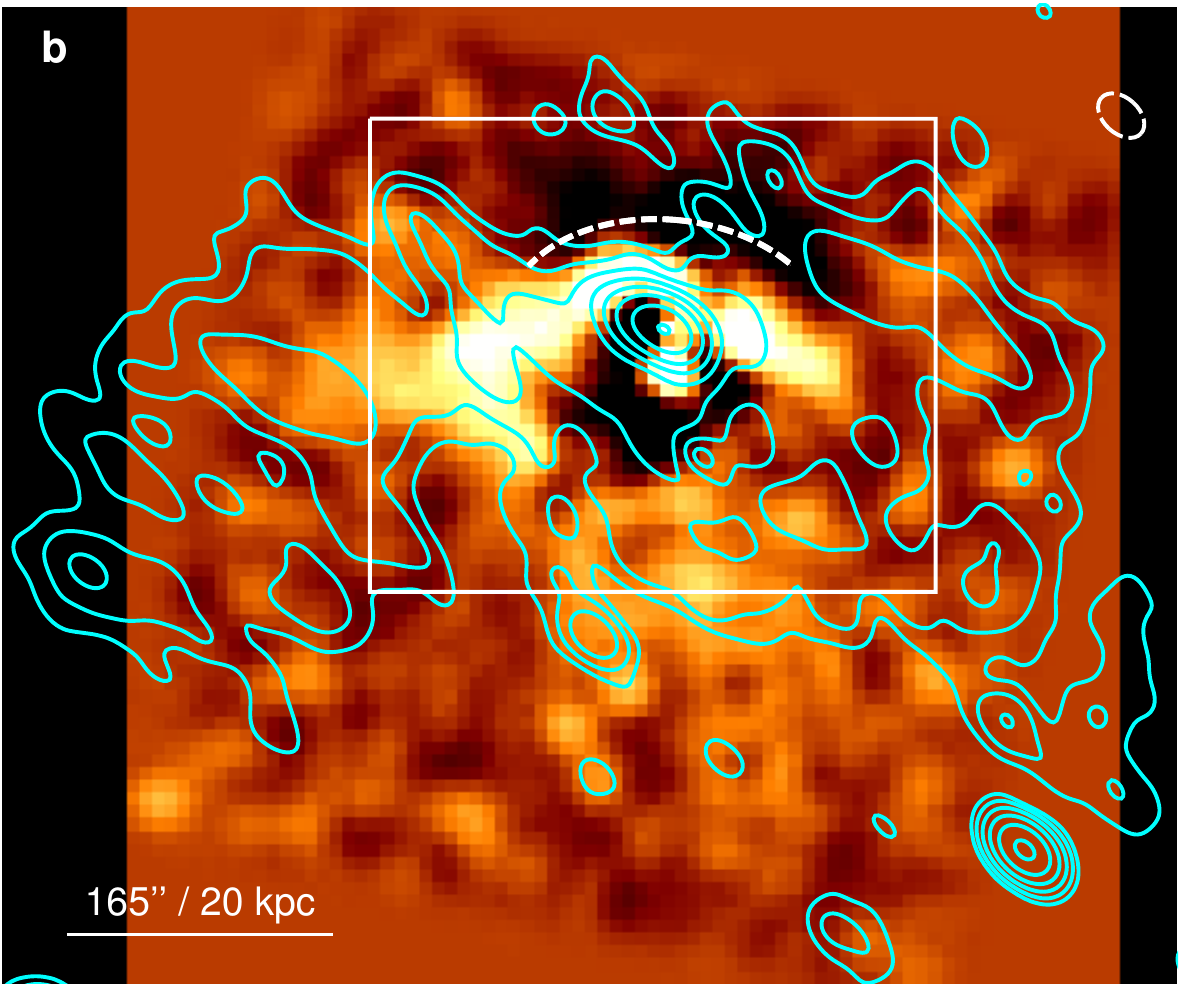}
\caption{{\em a}: 0.3--2 keV {\em Chandra} image of NGC\,1407,
Gaussian--smoothed with a 10$^{\prime\prime}$ kernel, with
overlaid the 610 MHz contours at the resolution of 
$5.6^{\prime\prime}\times4.3^{\prime\prime}$ from G11. 
Contours are spaced by a factor of 2 starting from $3\sigma$=0.3 mJy
beam$^{-1}$. The white arc marks the approximate position of the 
surface brightness edge, $\sim$70$^{\prime \prime}$ north of the
galaxy core. 
{\em b}: 0.3--2.0 keV {\em Chandra} surface brightness residuals after
fitting and subtracting two circular $\beta$-models and a flat model
for the background, binned by a factor 16 and smoothed with a
Gaussian with a radius of 3 pixel (23.66\arcsec). Contours at 330 MHz,
 at the resolution of 
$33^{\prime\prime}\times22^{\prime\prime}$, are overlaid starting at 1.5 mJy beam$^{-1}$
and then scaling by a factor of 2. The white box indicates the area of
panel {\em a}. In both panels the radio beam is indicated by the white
dashed ellipse.}
\label{fig:chandra}
\end{figure*}
%
%%%%%%%%%%%%% end of FIG 9 %%%%%%%%%%%%%%%%%%%%%%%%%%%%%%%%%
%

\section{X-ray Observations and Data Analysis}
\label{sec:Xobs}

NGC\,1407 was observed during \chandra\ cycle 1, on 2000 August 16, for
$\sim$49~ks (ObsID 791), with the ACIS-S instrument operating in very faint
telemetry mode. A summary of the \chandra\ mission and instrumentation can
be found in \citet{2002PASP..114....1W}. The data were reduced and analysed
using \textsc{ciao} 4.2 and CALDB 4.3 following techniques similar to those
described in \citet{2007ApJ...658..299O} and the \chandra\ analysis
threads\footnote{http://asc.harvard.edu/ciao/threads/index.html}. The first
half of the observation suffered from significant background flaring, and
the final cleaned exposure time was 30.9~ks.

Point sources were identified using the \textsc{wavdetect} task, with a
detection threshold of 1$\times$10$^{-6}$, chosen to ensure that the task
detects $\leq$1 false source in the ACIS-S field, working from a 0.3-7.0
keV image and exposure map. All point sources were excluded except 
the source corresponding to the AGN.  Spectra were extracted using
the \textsc{specextract} task. Spectral fitting was performed in
\textsc{xspec} 12.7.0e. Abundances were measured relative to the abundance
ratios of \citet{1998SSRv...85..161G}. A galactic hydrogen column of
0.0542$\times10^{22}$ \pcmsq\ and a redshift of 0.0059 were assumed in all
fits. Spectra were grouped to 20 counts per bin, and counts at energies
outside the range 0.5-7.0 keV were ignored during fitting.

Background spectra were drawn from the standard set of CTI-corrected ACIS
blank sky background events files in the \chandra\ CALDB. The exposure time
of each background events file was altered to produce the same 9.5-12.0 keV
count rate as that in the target observation. Very faint mode background
screening was applied to both source and background data sets.

\xmm\ has also observed NGC\,1407 (ObsId 0404750101), with an exposure of
$\sim$66~ks. A detailed summary of the \xmm\ mission and instrumentation
can be found in \citet[and references therein]{2001A&A...365L...1J}. The
observation is centred close to NGC\,1400, with NGC\,1407 $\sim$9 $^{\prime}$
off-axis.  To provide a comparison with \chandra\ image analysis, we
reduced this data using \textsc{sas} v10 following the methods described in
\citet{2011MNRAS.416.2916O}. The EPIC-MOS instruments were operated in full
frame and the EPIC-pn in extended full frame mode, with the thin optical
blocking filter.  Periods including background flaring, when the total
count rate deviated from the mean by more than 3$\sigma$, were excluded,
leaving useful exposures of $\sim$38~ks (MOS) and $\sim$26~ks (pn). Point
sources were identified using \textsc{edetect$\_$chain}, and regions
corresponding to the 85 per cent encircled energy radius of each source
(except that at the peak of the diffuse emission) were excluded.

\section{X-ray analysis}

\subsection{The X-ray images}\label{sec:xrayim}

We initially examined the structure of the hot intra-group medium (IGM)
using exposure-corrected 0.3-2~keV \chandra\ and \xmm-Newton images. Previous
\rosat\ observations have shown NGC\,1407 to be more X--ray luminous than
NGC\,1400, and that there is a clump of emission between the two
\citep[e.g.,][]{2003ApJS..145...39M}. Figure \ref{fig:xmm} shows the \xmm\ image,
adaptively smoothed using the \textsc{sas} \textsc{asmooth} task with a
signal-to-noise ratio of 10. The image confirms the features observed in
the \rosat\ data. The 240~MHz emission (contours) extends well beyond the brightest
part of the X--ray emission associated with NGC\,1407.

The clump of emission between NGC\,1407 and NGC\,1400 consists of both
point sources and diffuse emission. Although brighter to the east of
NGC\,1400, the diffuse emission extends across the galaxy and to the west as
far as the edge of the field of view. The origin of this trail is
unclear. NGC\,1400 is known to have a velocity offset from the group mean
by $\sim$1100 km s$^{-1}$ \citep{1994A&A...283..722Q}, which is
$\sim$3 times the velocity dispersion of the group ($\sigma_v \sim$
370 km s$^{-1}$) and $\sim$2 times the sound speed $v_{\rm sonic}$
(for a 1.2 keV gas, $v_{\rm sonic} \sim 520$ km s$^{-1}$). Therefore, the trail could be
tidally or ram-pressure stripped material. However, there is no clear 
connection with NGC\,1407.

Figure \ref{fig:chandra}a shows the \chandra\ image, smoothed with 
a 10\arcsec\ Gaussian. The
emission is asymmetric, being swept back from the northern side, forming
two wings to the east and west, and more extended to the south. There is
some indication of an edge, or drop in surface brightness, on the
northern side. To investigate this structure further, we modelled and
subtracted the mean surface brightness distribution to reveal any residual
structures. Two $\beta$-models were used to represent the galaxy and group
emission, and a flat model for the background. Fitting was carried out in
\textsc{ciao} \textsc{sherpa}, correcting for exposure using a
monoenergetic 0.85~keV exposure map (the energy was chosen to match the peak
photon energy of the data). After subtraction of the model, the residual
map is smoothed to bring out any
remaining structures. The result is shown
in Figure \ref{fig:chandra}b. An arc of positive residuals extends
across the northern side of the galaxy, with strong negative residuals 
to its north and south. The arc is narrow in the centre, and broader
in the wings, particularly on the east side. This again suggests the 
presence of a surface brightness edge
on the north side of the galaxy, with denser or cooler gas forming the
bright arc, and the northern negative residual indicating the sharp drop
seen in Figure \ref{fig:chandra}a. The negative residual in the galaxy 
core is likely caused by the bright arc and central point sources
biasing the inner part of the model to high values. We note that
using an elliptical beta model does not result in any significant
changes to the structures but fails to model the core and central point 
source properly.

%
%%%%%%%%%%%%%%%%%%%%% Fig. 10 SB profile %%%%%%%%%%%%%%%%
%
\begin{figure}
\includegraphics[width=\columnwidth]{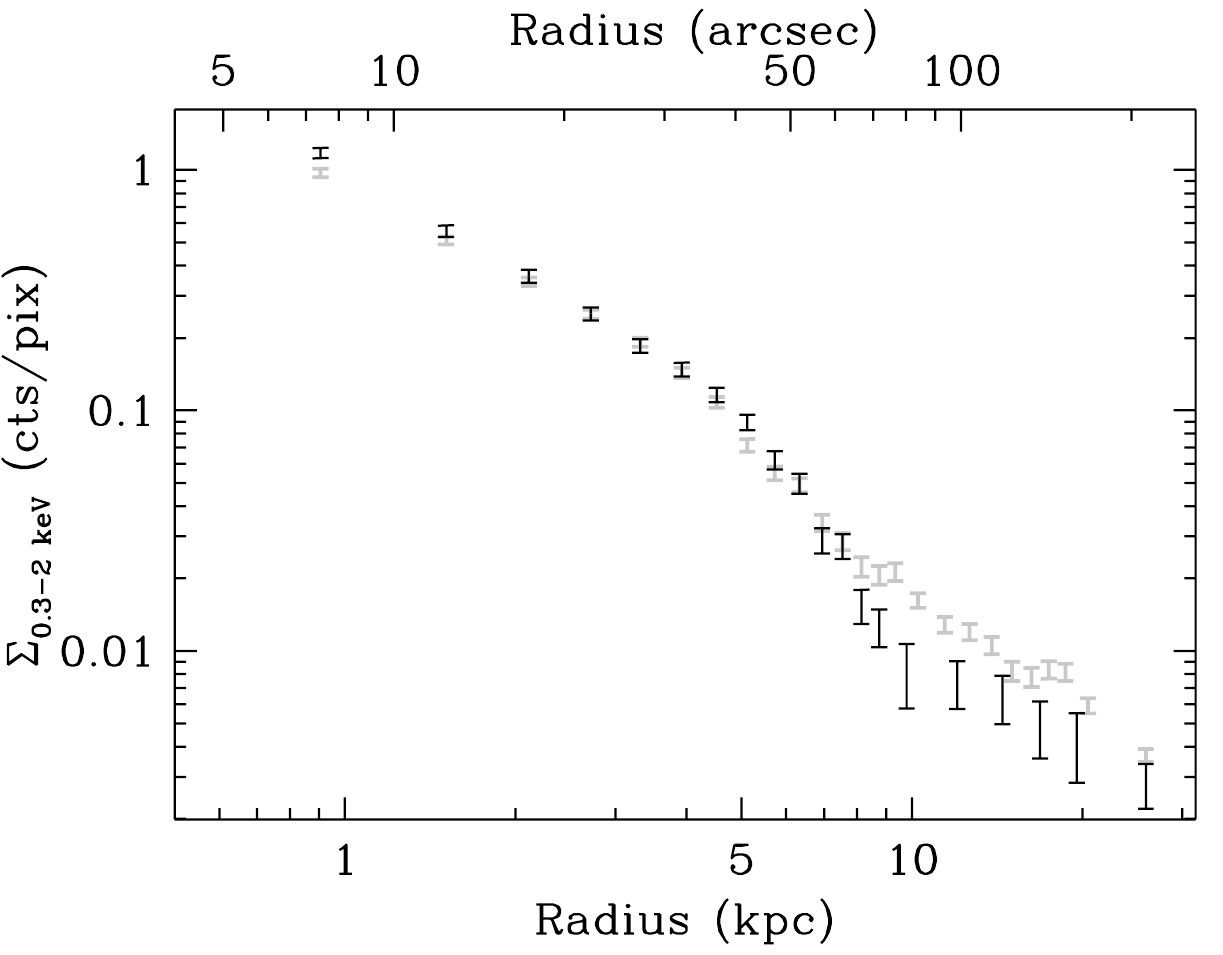}
\caption{\label{fig:XSB}Exposure-corrected 0.3-2~keV surface
  brightness in a 125\deg\ 
arc northward (black) and averaged over the remainder of the galaxy
(grey). Error bars 
indicate 1$\sigma$ uncertainty.}
\end{figure}
%
%%%%%%%%%%%%%%%%%%% End of Fig. 10 %%%%%%%%%%%%%%%%%%%%
%

Figure~\ref{fig:XSB} shows the 0.3-2~keV exposure--corrected surface
brightness profile extracted across the northern arc, using a 125\deg\
sector, with the mean surface brightness profile across the remainder of
the galaxy for comparison. The surface brightness drops further in the north
than in the rest of the halo, and the profile first steepens (between
$\sim$5-10~kpc), then flattens outside $\sim$10~kpc. There is no
  clear discontinuity in the profile, which would indicate the presence of
  a shock or cold front. However, the exposure may simply be too shallow
to detect such a feature; relatively wide spatial bins are required to
determine reliable surface brightness values.

\subsection{Gas properties}
\label{sec:Xgas}

\begin{figure}
\includegraphics[width=\columnwidth]{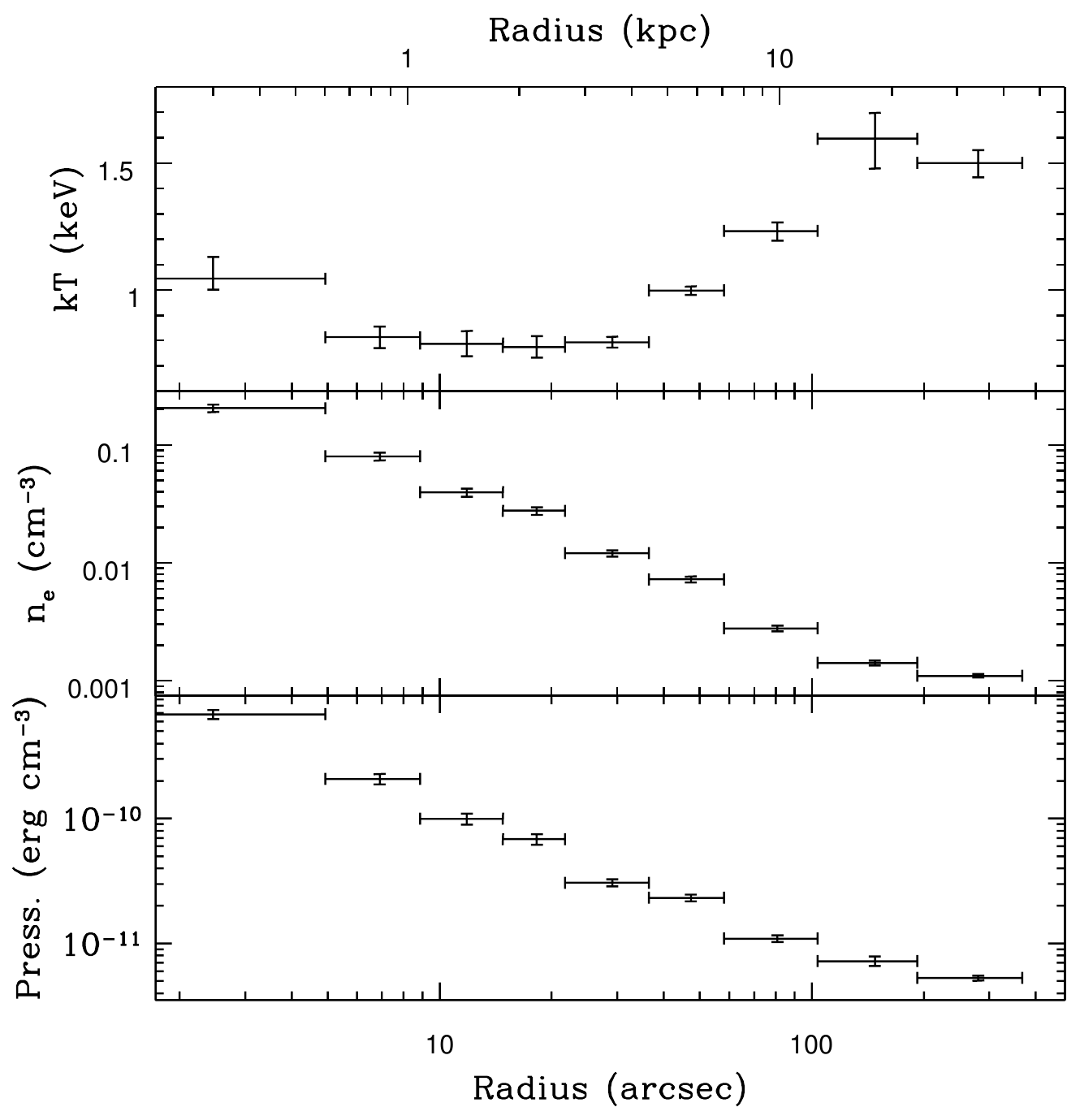}
\caption{\label{fig:press} Radial profiles of temperature, density and
  pressure in NGC\,1407. Density and pressure 
in the outermost bin may be overestimated since it includes 
emission from the outer parts of the IGM along the line of sight.}
\end{figure}

To determine the physical properties of the IGM, we extracted a radial
spectral profile centred on the galaxy. Since both the gas around 
NGC\,1407 and the larger-scale group emission show signs of disturbed, 
asymmetric structures, we chose to use simple circular annuli
containing 1000-3000 net counts. The radio emission is observed throughout
the core and over much of the \chandra\ field of view, therefore 
we do not exclude the corresponding region, since to do so would restrict the profile
  to the outer parts of the group. However, we note that if cavities are
present coincident with the radio emission, this will affect the results of
the spectral fits, both in terms of density (owing to the partial filling
factor of the X--ray emitting plasma) and temperature (since the
distribution of temperatures along the line of sight will change). The
spectra were fitted with a deprojected, absorbed APEC model. Abundances
were tied between bins where necessary to stabilise the deprojection.
The typical abundance was $\sim$0.6~Z$_\odot$.
Figure~\ref{fig:press} shows the resulting temperature and density
profiles, and the pressure profile derived from them. Pressure was
calculated as $P=nkT$ where $n=2n_e$.

%
%%%%%%%%%%%%%%%%%% Fig. 11 - KT %%%%%%%%%%%%%%%%%%%%%%%%
%
%
%%%%%%%%%%%%%%%% End of Fig. 11 %%%%%%%%%%%%%%%%%%%%%%%%%%
%

The temperature profile is in reasonable agreement with previous spectral
studies \citep[e.g.,][]{2007MNRAS.380.1554R,2008ApJ...687..986D}, showing a cool
core surrounded by hotter intra-group emission. Both density and pressure
profiles follow unexceptional approximate powerlaws.

\subsection{X-ray temperature map}

To examine the spatial variation of temperature in the gas, we prepared a
temperature map using the technique developed by
\cite{2009ApJ...705..624D} which
takes advantage of the close correlation between the strength of lines in
the Fe-L complex and gas temperature in $\sim$1~keV plasma.  Most of the
emission from such gas arises from the L--shell lines from Fe-XIX
(Ne--like) to Fe-XXIV (He--like). For CCD resolution spectra, these lines
are blended to form a single broad peak between approximately 0.7 and 1.2
keV. The centroid or mean photon energy of this peak increases with the
temperature of the gas as the dominant ionisation state of Fe shifts from
Fe XIX in 0.5~keV gas to Fe XXIV in 1.2 keV gas, with Fe XVII and Fe XVIII
providing the strongest line emission at the temperatures seen in the core
of NGC~1407. The mean photon energy of the blended L--shell lines is
independent of energy above $\sim$1.3~keV.

We can thus estimate the temperature distribution of the gas by mapping the
mean photon energy in the 0.7-1.2~keV band. We estimated the mean gas
properties from the spectral fits described in Section~\ref{sec:Xgas},
taking an abundance 0.6 $Z_{\odot}$ as representative and simulated spectra for a
range of temperatures, with redshift set to that of NGC\,1407 and the
hydrogen column set to the Galactic value.  These showed that for
temperatures between $\sim$0.5 and $\sim$1.3~keV, the relationship is
approximately linear ($kT=-6.38+7.83<E>$). Prior testing in other systems
suggests that the relationship is relatively insensitive to variations in
abundance and hydrogen column \citep{2009ApJ...701.1560O}.

\begin{figure}[ht!]
\epsscale{1}
\plotone{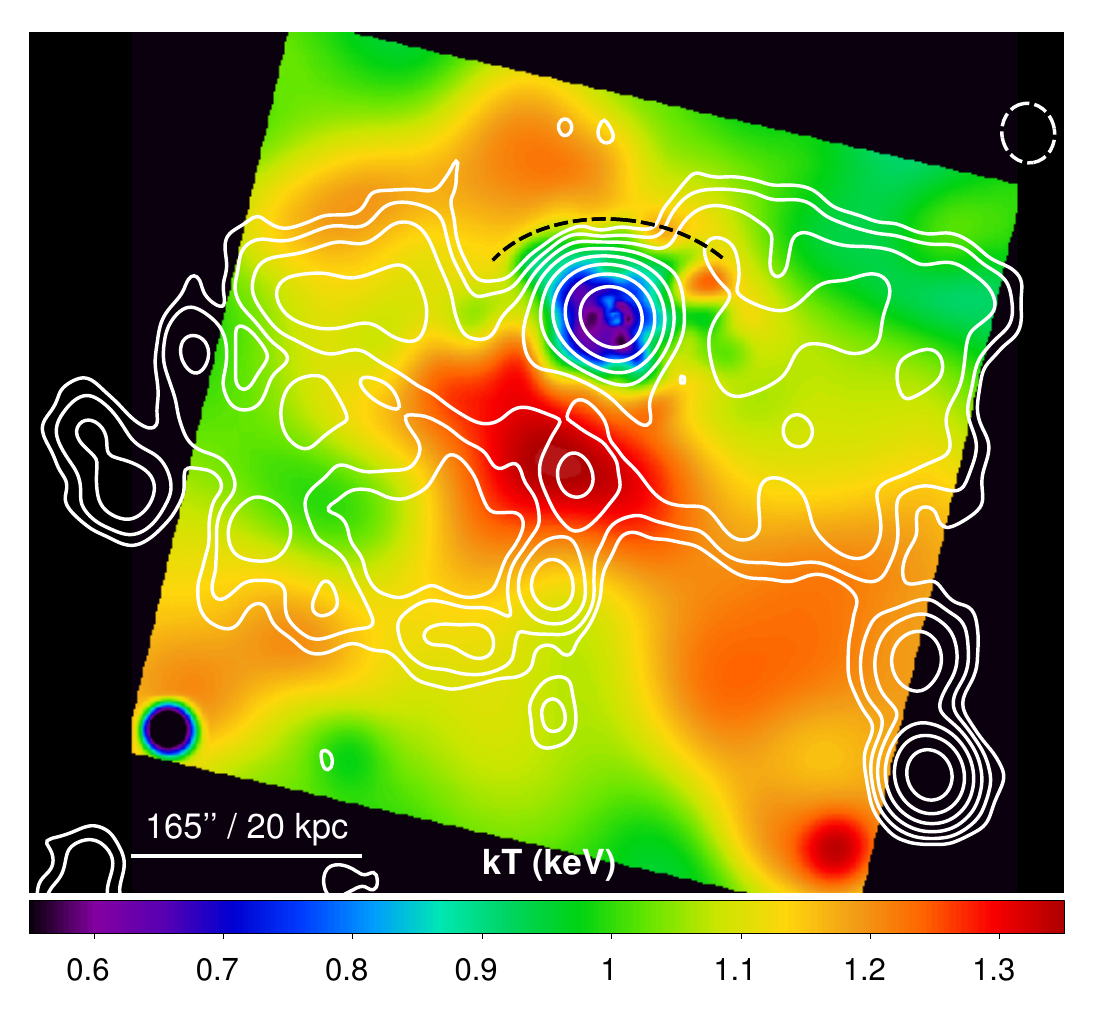}
\caption{{\em Chandra} adaptively smoothed Fe-peak temperature
 map, with the low resolution 610 MHz contours overlaid (same as
 Fig.~\ref{fig:610lr}{\em a}). The black dashed arc marks the surface 
brightness edge in Fig.~\ref{fig:chandra}. The radio beam is indicated by the white
dashed ellipse. }
\label{fig:temp}
\end{figure}

Using this technique, we produced the map shown in Figure \ref{fig:temp}.
The temperature in the galaxy core is 0.5-0.7~keV, rising to 1.2-1.3~keV
north and south of the galaxy. We note that the highest temperatures in the
map are somewhat lower than those in the radial profile
(Fig.~\ref{fig:press}), since the mapping technique becomes insensitive to
temperature increases above $\sim$1.3~keV.  The cool core region is
relatively circular. The black dashed arc marks the location of the X-ray
surface brightness edge.  The gas behind the edge seems to be
  cooler ($\sim$0.9-1~keV) than the gas right in front ($\sim$1.2~keV),
  suggesting that the feature may be a cold front. However, the exposure is
  too shallow to allow temperature measurements on a scale comparable to
  the size of the feature. We therefore cannot rule out the alternative
  possibilities that there is no strong discontinuity, or that a weak shock
  is present. A deeper \textit{Chandra} pointing would be required to
  constrain the gas temperature behavior across the edge and determine the
  real nature of this feature.

The cool ($\sim$1~keV) regions immediately east and west of the galaxy core
correspond with the wings of the arc of excess surface brightness seen in
Figure~9, showing that it is caused at least in part by the increased
emissivity of this cooler material. This may indicate that material has
been drawn out of the galaxy core and either heated or
mixed with warmer IGM gas.

%
%%%%%%%%%%%%%% FIG 12 - Tmap %%%%%%%%%%%%%%%%%
%
%
%%%%%%%%%%%%% End of Fig. 12%%%%%%%%%%%%%%%%%%%%%%%%%%%%%%%%%%
%

\section{Discussion}
\label{sec:discussion}

\subsection{Restarted activity in NGC\,1407}\label{sec:age}

The radio images presented in Section \ref{sec:images} show
faint, diffuse emission, approximately 80 kpc across, 
which wholly encloses and dwarfs a small-scale ($\sim8$ kpc) 
double source at the center of NGC\,1407. 
Based on former \gmrt\ observations, 
G11 suggested that the diffuse outer emission was
produced during an earlier cycle of activity of NGC\,1407. 
The spectral analysis presented in this paper corroborates the multiple 
outburst scenario. The large-scale emission is found to have 
an ultra-steep radio spectrum, with $\alpha=1.82$, as typically 
observed for highly evolved and fading radio sources \citep[e.g.,][]{
1994A&A...285...27K, 2007A&A...470..875P, 2007A&A...476...99G, 2011A&A...526A.148M}.
The spectral age analysis suggests that the radio plasma in such component
is at least $\sim$300 Myr old. Furthermore, the radiative model
that better describes the spectrum of this emission requires a 
switch-off of the nuclear engine $\sim$170-190 Myr ago, 
followed by a dying phase. The inner double has instead a 
spectral index of $\alpha=0.7$ and a radiative age of $\sim 30$ Myr, 
consistently with being a currently active and relatively young radio
source.

The multi-scale radio morphology, combined with the 
distinct spectral properties of its components, makes NGC\,1407
another remarkable example of a {\em nesting} radio galaxy, other nearby
examples being, for instance, 4C29.30 \citep{2007MNRAS.378..581J}
and Hercules A \citep{2005MNRAS.358.1061G}. 
As double-double radio galaxies, these peculiar radio sources provide 
evidence for recurrent radio activity in elliptical galaxies. In the
specific case of NGC\,1407, the small, young double source is
currently fed by the central AGN, while the
diffuse, steep-spectrum component is associated with relic plasma,
which was injected during an earlier radio outburst of the AGN
occurred at least $\sim$300 Myr ago. 

It is important to note that our radiative ages have been estimated
neglecting adiabatic expansion. This could be a reasonable
approximation for a relaxed, aged plasma in the dying phase such
as that in the large-scale component. However, expansion losses 
can be important for the inner double, as well as for the earlier, 
active phase of the large-scale component. Indeed, neglecting their 
effect may lead to an underestimate of the true source age.

Our age derivation is also based on the assumption of 
uniform and constant magnetic field across the source. A 
calculation of the effects of magnetic field evolution is 
beyond the purpose of the present paper, and we refer 
to \cite{1994ApJS...90..955R}, \cite{1999ApJ...512..105J}, 
\cite{2000AJ....119.1111B}, and references therein, for detailed works on  
the implications of magnetic field evolution for the source aging.
In the case of very aged emission as that in 
NGC\,1407, the relativistic plasma may be partially mixed 
with the hot thermal gas. In this case, magnetized filaments 
can be produced in the radio volume due to plasma instabilities 
and the total synchrotron emission would result from the convolution 
of different spectra produced by relativistic electrons emitting in 
regions with different magnetic field strength
\citep[e.g.,][]{2004ApJ...601..778T}.
The resulting convolution of the synchrotron
kernel with magnetic field intensity and geometry yields a total
spectrum which is stretched and thus not straightforwardly related to
the spectrum of the emitting electrons
\citep[e.g.,][]{1996ApJ...457..150E,1997ApJ...488..146K}.
In this case, 
standard aging analyses, based on the position of the 
break frequency in the synchrotron spectrum, can give 
incorrect estimates of the age of the radio emitting electrons.

\subsection{Energy output of the radio outbursts}

We compared the energy output associated with synchrotron radiation
from the two radio outbursts in NGC\,1407. Using the radio luminosity 
at 240 MHz and $\alpha_{\rm
  inj}$ in Table 5, we calculated the total radio luminosity over the
frequency range 10 MHz--100 GHz adopting the expression in \cite{1987ApJ...316...95O}. We found a bolometric radiative power of
$L_{\rm tot}  \sim 1.6 \times 10^{42}$ erg s$^{-1}$ for the inner 
double and $L_{\rm tot}  \sim 3.2 \times 10^{42}$ erg s$^{-1}$ for the 
large-scale emission.
Based on our estimates of the radiative age of the two components
(Table 5), 
we found that the current outburst has released an energy of $E_{\rm tot,
  \, syn} \sim 1.5 \times 10^{57}$ erg so far, while the energy of the
former episode of activity, $E_{\rm tot, \, syn} \sim 2.6-3.2\times
10^{58}$ erg, is one order of magnitude larger. 

It is well known that the synchrotron luminosity is an underestimate 
of the total energy output of a radio source, which is believed to 
be dominated by the mechanical work done by the radio jets on the 
external medium. The mechanical jet power of radio sources can be inferred, for
instance, in those groups/clusters with depressions (cavities) in their 
X-ray surface brightness, interpreted as rising bubbles of 
relativistic plasma inflated by the central AGN 
\citep[e.g.,][]{2004ApJ...607..800B,2008ApJ...686..859B,2010ApJ...720.1066C,2011MNRAS.416.2916O}.
Here, it is estimated that the ratio of the mechanical 
to synchrotron luminosities is a few to few thousands for
powerful radio sources, and up to several thousand for weaker sources
\citep[e.g.,][]{2004ApJ...607..800B}. In those special systems with multiple cavities, 
interpreted as signature of repeated AGN outbursts, it is also possible to
compare the energy outputs of the different episodes of activity.
In some systems, such as Hydra A \citep{2007ApJ...659.1153W}
and NGC\,5813 \citep{2011ApJ...726...86R}, it is found that the past epoch of
activity is the most energetic, suggesting that the mean jet power
changed significantly over time or that the current outburst is still ongoing.
In other systems (e.g., A\,262, \cite{2009ApJ...697.1481C}), there is 
evidence for an opposite outburst trend, which may
reflect an increase of the fueling of the AGN with time.

The current X-ray images of NGC\,1407 do not show evidence of
X-ray cavities associated with the large-scale diffuse radio
structure. \cite{2010ApJ...712..883D} report a possible small
(0.7$\times$0.4~kpc) cavity in the group core, but its identification
is dependent on the image processing employed and it appears to be 
uncorrelated with the small-scale active radio source. 
A direct measurement of the mechanical power of the radio jets 
in either period of activity is therefore not possible. If the mean
jet mechanical power to radio power ratio has remained constant
over the outburst history, NGC\,1407 would then be another 
system in which the total energy output from the AGN is decreasing
with time. On the other hand, the most recent episode of activity 
may be still ongoing (its radiative age is $\sim$30 Myr), and it is
possible that the jet power may increase.

\subsection{Pressure comparison}

Assuming energy equipartition arguments, we can derive the 
non-thermal pressure in the radio source and compare it to the 
pressure of the surrounding X-ray gas. Under the assumptions 
listed at the beginning of Sect.~\ref{sec:spectra} (see point 3), and
adopting $\alpha_{\rm inj}$ and $B_{\rm eq}$ in Table 5, we calculated
a radio pressure in the large-scale component of $P_{\rm radio, diffuse}= 2.5
\times 10^{-13}$ erg cm$^{-3}$ and $P_{\rm radio, S3}= 2.8
\times 10^{-12}$ erg cm$^{-3}$ for the inner double. 

Even though the X-ray pressure profile in Fig.~\ref{fig:press} does not
cover the whole extent of the radio emission, it is clear that the
pressure of the X-ray gas is at least one order of magnitude 
higher than the radio pressure, with values ranging from 
$\sim 10^{-11}$ erg cm$^{-3}$ at $\sim$10 kpc 
from the center to $\sim5 \times 10^{-12}$ erg cm$^{-3}$ at $\sim 40$ kpc.
This is not unusual for cool-core systems, where a similar
pressure discrepancy is often found \citep[e.g.,][]{2005MNRAS.364.1343D},
suggesting a departure from equipartition conditions or an additional 
pressure support in the radio lobes, for instance, in the form of
thermal gas \citep[e.g.,][]{2010MNRAS.407..321O}. Alternatively, an
energetically dominant population of non-radiating relativistic particles
(protons) may contribute to the internal pressure. The achievement of
pressure balance in NGC\,1407 would then require a ratio of energy 
in non-radiating particles to the energy in electrons $k \sim
100-1000$. These values are in the range typically 
found for samples of centrally-located radio galaxies in groups and clusters
\citep{2004MNRAS.355..862D,2008ApJ...686..859B,2008MNRAS.386.1709C,2010ApJ...714..758G}
and suggest that that the lobes have been fed by {\em heavy} 
jets or have entrained thermal material during their propagation
through the cluster/group atmosphere \citep[e.g.,][]{2010MNRAS.407..321O}.
The diffuse and distorted
morphology of the large-scale emission in NGC\,1407, combined 
with the very long age inferred for this component, suggest that
a mixing of the radio plasma and ambient thermal gas has already
occurred at some level, favoring the entrainment scenario.

\subsection{Motion of the galaxy}

The morphology of NGC\,1407 in the X-ray band (Fig.~\ref{fig:xmm} and
Fig.\ref{fig:chandra}) suggests that the galaxy is in motion. A surface
brightness edge is visible in the north side of the galaxy with gas swept
back to the east and west into two cool wings.  There is some indication of
a possible discontinuity in the surface brightness profile of the northern
quadrant of the galaxy at this position (Fig.~10). If confirmed, this
feature would suggest the presence of a shock or a cold front. The Fe-peak
temperature map shows variations across the edge which seem consistent with
the the hypothesis of a cold front, i.e., with the cooler gas behind the
front (Fig.~12). However, the available data are not deep enough to
constrain the temperature jump across the edge and thus the nature of this
feature.

Overall, NGC\,1407 appears to be moving northward with its halo being
stripped by the surrounding IGM. The location of the putative front is
consistent with this scenario. If the edge is a cold front, it may
indicate that the gas is sloshing in response to a recent 
interaction, such as a close passage, of NGC\,1400. The X-ray trail
visible in the \xmm\ image (Fig.~8) could originate from 
the same interaction. 

The diffuse radio structure also shows a swept-back shape, but 
on a larger scale, similar to wide-angle tail (WAT) radio galaxies,
although no jets are present. As discussed above, this emission is
found to be the remnant of a former radio outburst of the central
galaxy, consistent with the absence of jets. The X-ray wings seem 
to be inside the radio emission, and the radio contours appear
compressed at the northern edge (e.g., Fig.\ref{fig:temp}).

The similarities in the radio and X-ray morphologies suggest that the 
two structures are co-spatial and both affected by the motion
of the galaxy. The X-ray wings may then represent a wake of cooler
galactic material which has been stripped by the same ram-pressure
that bent the diffuse radio emission. 

There are several ways in which we can obtain rough 
estimates of the velocity of NGC\,1407 relative
to the IGM. If we assume, for simplicity, that the bending of the 
diffuse radio emission is caused by ram pressure, we can use the 
Euler equation in the form

$$\frac{\rho_{\rm radio} v_{\rm radio}^2}{r_c} = \frac{\rho_{\rm IGM}
  v_{\rm gal}^2}{r_{\rm radio}}$$

\citep[e.g.,][]{1985ApJ...295...80O}, where 
$r_c$ is their curvature radius, $r_{\rm radio}$ is the 
radius of the radio {\em tails}, $\rho_{\rm radio}$ and 
$v_{\rm radio}$ are the density and velocity of the radio-emitting plasma, 
$v_{\rm gal}$ is the velocity of the galaxy relative to the IGM, and
$\rho_{\rm IGM}$ is the density of the IGM.
From the images presented in Sect.~\ref{sec:images}, we estimate
$r_{\rm radio} \sim 10$ kpc and $r_c \sim 30$ kpc. Based on the
spatial extent of the diffuse radio emission (LLS$\sim 80$ kpc) and age 
of the active phase $t_{\rm CI} \sim 90-130$ Myr (Table 5), 
we derive a first order estimate of the growth velocity of the radio
source of $v_{\rm radio} = {\rm LLS}/t_{\rm CI} \sim 0.002c-0.003c$, 
where $c$ is the speed of light. Finally, from the \chandra\ data, 
we obtain $\rho_{\rm IGM} \sim 0.0015$ cm$^{-3}$ within the central 
50 kpc and assume $\rho_{\rm radio} \sim 10^{-3} \rho_{\rm IGM}$
\citep[e.g.,][]{2011ApJ...743..199D}. This gives a very low
  velocity, only $v_{\rm  gal} \sim$ 20 km s$^{-1}$. 

Alternatively, we can make a simple calculation assuming that the
galaxy is moving north, that the radio outburst was 
located where we now see the southern boundary of the diffuse 
structure, and that it has moved at a constant velocity since 
that time. Neglecting projection effects and basing our estimate 
on the 240~MHz map, the galaxy would then have traveled 
28-35~kpc over a period of $\sim$300~Myr, 
suggesting a velocity of 90-115 km s$^{-1}$. 

These velocity estimates conflict to some extent with the X-ray
  morphology. If a cold front is present in NGC~1407, we would expect the
  galaxy to be moving at a significant fraction of the sound speed. For a
  temperature of 1.1~keV, as is found immediately north of the surface
  brightness edge, the sound speed is $v_{sonic}\sim$500 km s$^{-1}$, so a
  velocity of at least 250 km s$^{-1}$ would be likely.

However, there are large uncertainties on our estimates. If
  sloshing is taking place, the most likely cause is perturbation by a
  tidal encounter with NGC\,1400. Given the 1100 km s$^{-1}$ velocity offset
  between NGC\,1400 and the group mean, it is clear that sloshing motions
  would include a significant line of sight component. Since we see signs
  of a surface brightness edge in the X-ray and compressed contours in the
  radio, the velocity in the plane of the sky is greater than or comparable
  to that in the line of sight, but this still leaves an uncertainty of up
  to a factor $\sqrt{2}$ in the true velocity of NGC\,1407. The velocity
  could thus be as high as 160 km s$^{-1}$, or 0.3$v_{sonic}$.

Projection effects could have an impact on our estimate of the
  ram-pressure timescale. If the diffuse structure is in fact two old radio
  lobes, the axis of the jet which formed them may have been at an angle to
  the line of sight. In this case the lobes would be at a larger radius
  than we have assumed, and would experience lower external pressures.
  Mixing of entrained thermal plasma into the lobes would raise their
  effective density. Both of these factors could increase the ram--pressure
  velocity estimate, though it is difficult to place limits on their
  impact.

As discussed above, the X-ray images of NGC\,1407 suggest that the
  group is not relaxed. A connection between bending of radio jets and
  sloshing induced by minor mergers has been proposed in cluster cores
  \citep[e.g.,][]{2004ApJ...616..178C,2006ApJ...650..102A,2012arXiv1203.2312M},
  and it seems possible that NGC\,1407 is an example of this process at work
  in a galaxy group.  In particular, we speculate that the group core is
  sloshing along the North-South axis, in response to a possible recent
  interaction with the nearby group galaxy NGC\,1400. The gas motions
  induced by such sloshing may be then shaping both the X-ray and radio
  structures. There is a degree of tension between the velocities estimated
  from the radio and X-ray data. Resolving this issue would require deeper
  X--ray data, capable of detecting any cold front and providing a more
  reliable velocity measurement, and of determining the filling factor of
  the diffuse radio structure.

\section{Summary}

We have examined the complex radio emission associated with the galaxy
NGC\,1407, at the center of the group Eridanus A. In a previous work,
we have suggested that the galaxy experienced two distinct radio
outbursts. Thanks to new, deep \gmrt\ observations at 240 MHz, 330
MHz and 610 MHz, combined with archival \vla\ data, we confirm the
multiple-outburst scenario. NGC\,1407 appears to be currently active
in the radio band in the form of a small-scale ($\sim$ 8 kpc) double
source associated with the galaxy. The analysis of the
integrated radio spectrum indicates that the source has a spectral
index of 0.7 and a radiative age of $\sim$30 Myr, consistent
with being an active and relatively young radio source. 
The double is wholly embedded in a faint, diffuse component, 
extending on a scale which is almost an order of magnitude larger 
($\sim$80 kpc). The large-scale emission is found to have an
ultra-steep spectrum, with $\alpha=1.8$, as typically found for
highly evolved and/or dying radio galaxies. The spectral age analysis
shows that the radio plasma in such a component is at least $\sim$300 Myr old, 
supporting the idea that it was generated during a former radio
outburst of NGC\,1047. We estimated that the radio nucleus switched
off $\sim$170-190 Myr ago, and after this time the source entered a
dying phase. 

We compared the synchrotron energy output associated with the 
two radio outbursts and found that the energy released during the 
duration of its current activity is nearly an
order of magnitude lower than that from the former outburst. 
Numerous studies in the literature have shown
that the synchrotron luminosity is an underestimate of the total
energy output of a radio source, which is dominated by the mechanical
work done by the jets on the external medium. If the mean jet power
remained constant over the whole outburst history of NGC\,1407, then 
NGC\,1407 would be another example of a system in which the energy
output is decreasing with time, other examples being Hydra A and
NGC\,5813. On the other hand, the most recent episode of activity may
be still ongoing, and it is possible that the jet power may increase.

We analyzed {\em XMM-Newton} and {\em Chandra} observations of the group.
The diffuse emission surrounding NGC~1407 appears swept back to the south,
with wings extending to the east and west, and a possible surface
brightness discontinuity on the northern side of the galaxy.  Temperature
mapping shows the emission inside the edge to be cooler than that to the
north, but the data are shallow and the smoothing scale is too large to
confirm whether a cold front or weak shock is present.  The northern edge
coincides with regions of compressed radio contours, and the large--scale
radio structure also has a swept--back morphology, indicating that the
radio and X-ray structures are co--spatial and affected by the same forces.
In general, the radio and X--ray morphology are suggestive of northward
motion. We speculate that the group core may be sloshing on a north--south
axis within the larger group potential, having been tidally disturbed by an
encounter with the nearby elliptical NGC\,1400. Our estimated
  velocity in the plane of the sky is relatively low ($\sim$100 km
  s$^{-1}$), but given the large uncertainties it seems possible that the
  true velocity could be a significant fraction of the sound speed, in
  which case sloshing motions could produce the radio and X-ray features we
  observe.

Comparison of pressure estimates for the large--scale radio 
structure and the surrounding intra--group medium show the usual 
order--of--magnitude difference, the X--ray pressure estimate being
much higher than that found from the radio data. This may indicate a large
population of non--radiating particles within the radio structure,
with a total energy $\sim$100-1000 times that of the relativistic electron
population. If, as seems likely, the structure was formed by a previous
AGN outburst, these particles could be relativistic protons
accelerated within the original radio jets, or a thermal plasma
component which has been entrained during the propagation and expansion of the
jets and lobes. The latter possibility provides an explanation 
for the continued presence of the radio structure in the group core,
since buoyant forces, which would normally lift the lobes into the
outskirts of the group halo, will be suppressed if the lobes contain 
a large fraction of thermal plasma.

\acknowledgements
%\section{Acknowledgments}

We thank the staff of the GMRT for their help during the
observations. GMRT is run by the National Centre for Radio Astrophysics of
the Tata Institute of Fundamental Research. SG
acknowledges the support of NASA through Einstein Postdoctoral
Fellowship PF0-110071 awarded by the {\em Chandra}
X-ray Center (CXC), which is operated by the Smithsonian Astrophysical
Observatory (SAO). EOS acknowledges the support of the European Community
under the Marie Curie Research Training Network. 
Basic research in astronomy at the Naval Research Laboratory is funded
by 6.1 Base funding. JMV, LPD and EOS
acknowledges the support of the Smithsonian Institution and CXC.
The National Radio Astronomy Observatory is a facility of the National Science 
Foundation operated under cooperative agreement by Associated Universities, Inc.
This research has made use of the NASA/IPAC Extragalactic
Database (NED) which is operated by the Jet Propulsion Laboratory, California
Institute of Technology, under contract with the National Aeronautics and Space
Administration.


\begin{thebibliography}{}



\bibitem[Ascasibar 
\& Markevitch(2006)]{2006ApJ...650..102A} Ascasibar, Y., \&
Markevitch, M.\ 2006, \apj, 650, 102 

\bibitem[Baars et 
al.(1977)]{1977A&A....61...99B} Baars, J.~W.~M., Genzel, R., Pauliny-Toth, I.~I.~K., \& Witzel, A.\ 1977, \aap, 61, 99 

\bibitem[Beck 
\& Krause(2005)]{2005AN....326..414B} Beck, R., \& Krause, M.\ 2005, Astronomische Nachrichten, 326, 414 


\bibitem[B{\^i}rzan et al.(2004)]{2004ApJ...607..800B} B{\^i}rzan, L., 
Rafferty, D.~A., McNamara, B.~R., Wise, M.~W., 
\& Nulsen, P.~E.~J.\ 2004, \apj, 607, 800 

\bibitem[B{\^i}rzan et al.(2008)]{2008ApJ...686..859B} B{\^i}rzan, L., 
McNamara, B.~R., Nulsen, P.~E.~J., Carilli, C.~L., 
\& Wise, M.~W.\ 2008, \apj, 686, 859 



\bibitem[Blundell 
\& Rawlings(2000)]{2000AJ....119.1111B} Blundell, K.~M., \& Rawlings, S.\ 2000, \aj, 119, 1111 

\bibitem[Brough et al.(2006)]{2006MNRAS.369.1351B} Brough, S., Forbes, 
D.~A., Kilborn, V.~A., Couch, W., \& Colless, M.\ 2006, \mnras, 369, 1351 



\bibitem[Brunetti et 
al.(1997)]{1997A&A...325..898B} Brunetti, G., Setti, G., \& Comastri, A.\ 1997, \aap, 325, 898 


\bibitem[Burns et al.(1994)]{1994ApJ...423...94B} Burns, J.~O., Rhee, G., 
Owen, F.~N., \& Pinkney, J.\ 1994, \apj, 423, 94 


\bibitem[Cavagnolo et al.(2010)]{2010ApJ...720.1066C} Cavagnolo, K.~W., 
McNamara, B.~R., Nulsen, P.~E.~J., et al.\ 2010, \apj, 720, 1066 



\bibitem[Chandra et al.(2004)]{2004ApJ...612..974C} Chandra, P., Ray, A., 
\& Bhatnagar, S.\ 2004, \apj, 612, 974 


\bibitem[Clarke et al.(2004)]{2004ApJ...616..178C} Clarke, T.~E., Blanton, 
E.~L., \& Sarazin, C.~L.\ 2004, \apj, 616, 178 

\bibitem[Clarke \& Ennslin (2006)]{2006AJ....131.2900C} Clarke, T. E., Ensslin, T. A.\ 2006, \aj, 131, 2900


\bibitem[Clarke et al.(2009)]{2009ApJ...697.1481C} Clarke, T.~E., Blanton, 
E.~L., Sarazin, C.~L., et al.\ 2009, \apj, 697, 1481 



\bibitem[Cohen et al.(2007)]{2007AJ....134.1245C} Cohen, A.~S., Lane, 
W.~M., Cotton, W.~D., Kassim, N.~E., Lazio, T.~J.~W., Perley, R.~A., 
Condon, J.~J., \& Erickson, W.~C.\ 2007, \aj, 134, 1245 

\bibitem[Condon et al.(1998)]{1998, AJ, 115, 1693}
Condon, J. J., Cotton, W. D., Greisen, E. W., et al., 1998, AJ, 115, 1693


\bibitem[Croston et al.(2008)]{2008MNRAS.386.1709C} Croston, J.~H., 
Hardcastle, M.~J., Birkinshaw, M., Worrall, D.~M., 
\& Laing, R.~A.\ 2008, \mnras, 386, 1709 


\bibitem[David et al.(2009)]{2009ApJ...705..624D} David, L.~P., Jones, C., 
Forman, W., et al.\ 2009, \apj, 705, 624 

\bibitem[De Vaucouleurs et al. (1991)]{1991rc3..book.....D}
De Vaucouleurs, G., De Vaucouleurs, A., Corwin Jr., H.G., Buta, R. J.,
Paturel, G., Fouque, P., Third Reference Catalogue of Bright Galaxies,
Version 3.9


\bibitem[Diehl 
\& Statler(2008)]{2008ApJ...687..986D} Diehl, S., \& Statler, T.~S.\ 2008, \apj, 687, 986 

\bibitem[Dong et al. (2010)]{2010ApJ...712..883D} Dong, R., Rasmussen, J. \& Mulchaey, J.~S.\ 2010, \apj, 712, 883


\bibitem[Douglass et al.(2011)]{2011ApJ...743..199D} Douglass, E.~M., 
Blanton, E.~L., Clarke, T.~E., Randall, S.~W., 
\& Wing, J.~D.\ 2011, \apj, 743, 199 

\bibitem[Dunn 
\& Fabian(2004)]{2004MNRAS.355..862D} Dunn, R.~J.~H., \& Fabian, A.~C.\ 2004, \mnras, 355, 862 

\bibitem[Dunn et al.(2005)]{2005MNRAS.364.1343D} Dunn, R.~J.~H., Fabian, 
A.~C., \& Taylor, G.~B.\ 2005, \mnras, 364, 1343 


\bibitem[Eilek et al.(1984)]{1984ApJ...278...37E} Eilek, J.~A., Burns, 
J.~O., O'Dea, C.~P., \& Owen, F.~N.\ 1984, \apj, 278, 37 


\bibitem[Eilek 
\& Arendt(1996)]{1996ApJ...457..150E} Eilek, J.~A., \& Arendt, P.~N.\
1996, \apj, 457, 150 


\bibitem[Forbes et al.(2006)]{2006PASA...23...38F} Forbes, D.~A., Ponman, T., Pearce, F., et al. \ 2006, PASA , 23, 38

\bibitem[Giacintucci et 
al.(2007)]{2007A&A...476...99G} Giacintucci, S., Venturi, T., Murgia, M., et al.\ 2007, \aap, 476, 99 

\bibitem[Giacintucci et al. (2008)]{2008A&A...486..347G} Giacintucci, S., Venturi, T., Macario, G., et al.\ 2008, \aap, 486, 347 

\bibitem[Giacintucci et al.(2011)]{2011ApJ...732...95G} Giacintucci, S., et 
al.\ 2011, \apj, 732, 95 (G11)


\bibitem[Gizani 
\& Leahy(2003)]{2003MNRAS.342..399G} Gizani, N.~A.~B., \& Leahy, J.~P.\ 2003, \mnras, 342, 399 


\bibitem[Gitti et al.(2010)]{2010ApJ...714..758G} Gitti, M., O'Sullivan, 
E., Giacintucci, S., et al.\ 2010, \apj, 714, 758 

\bibitem[Gizani et al.(2005)]{2005MNRAS.358.1061G} Gizani, N.~A.~B., Cohen, 
A., \& Kassim, N.~E.\ 2005, \mnras, 358, 1061 


\bibitem[Gomez et al.(1997)]{1997ApJ...474..580G} Gomez, P.~L., Pinkney, 
J., Burns, J.~O., et al.\ 1997, \apj, 474, 580 

\bibitem[Grevesse 
\& Sauval(1998)]{1998SSRv...85..161G} Grevesse, N., \& Sauval, A.~J.\ 1998, \ssr, 85, 161 


\bibitem[Greisen et al. (2009)]{2009AJ....137.4718G} Greisen, Eric W., Spekkens, Kristine, van Moorsel, Gustaaf A.\ 2009, \aj, 137, 4718

\bibitem[Jaffe 
\& Perola(1973)]{1973A&A....26..423J} Jaffe, W.~J., \& Perola, G.~C.\ 1973, \aap, 26, 423

\bibitem[Jamrozy et al.(2007)]{2007MNRAS.378..581J} Jamrozy, M., Konar, C., 
Saikia, D.~J., Stawarz, {\L}., Mack, K.-H., 
\& Siemiginowska, A.\ 2007, \mnras, 378, 581 

\bibitem[Jansen et 
al.(2001)]{2001A&A...365L...1J} Jansen, F., Lumb, D., Altieri, B., et
al.\ 2001, \aap, 365, L1 

\bibitem[Jones et al.(1999)]{1999ApJ...512..105J} Jones, T.~W., Ryu, D., 
\& Engel, A.\ 1999, \apj, 512, 105 

\bibitem[Kardashev(1962)]{1962SvA.....6..317K} Kardashev, N.~S.\ 1962, 
\sovast, 6, 317 


\bibitem[Katz-Stone 
\& Rudnick(1997)]{1997ApJ...488..146K} Katz-Stone, D.~M., \& Rudnick, L.\ 1997, \apj, 488, 146 


\bibitem[Komissarov 
\& Gubanov(1994)]{1994A&A...285...27K} Komissarov, S.~S., \& Gubanov, A.~G.\ 1994, \aap, 285, 27 

\bibitem[Lara et 
al.(1999)]{1999A&A...348..699L} Lara, L., M{\'a}rquez, I., Cotton, W.~D., Feretti, L., Giovannini, G., Marcaide, J.~M., \& Venturi, T.\ 1999, \aap, 348, 699 

\bibitem[Leahy et al.(1986)]{1986MNRAS.222..753L} Leahy, J.~P., Pooley, 
G.~G., \& Riley, J.~M.\ 1986, \mnras, 222, 753 


\bibitem[Mendygral et al.(2012)]{2012arXiv1203.2312M} Mendygral, P., Jones, 
T., \& Dolag, K.\ 2012, arXiv:1203.2312 


\bibitem[Myers 
\& Spangler(1985)]{1985ApJ...291...52M} Myers, S.~T., \& Spangler, S.~R.\ 1985, \apj, 291, 52 

\bibitem[Mulchaey et al.(2003)]{2003ApJS..145...39M} Mulchaey, J.~S., 
Davis, D.~S., Mushotzky, R.~F., \& Burstein, D.\ 2003, \apjs, 145, 39 

%\bibitem[2001]{murgia01} Murgia, M., 2001, PhD thesis, University of Bologna

\bibitem[Murgia et 
al.(2011)]{2011A&A...526A.148M} Murgia, M., et al.\ 2011, \aap, 526,
A148 


\bibitem[O'Dea(1985)]{1985ApJ...295...80O} O'Dea, C.~P.\ 1985, \apj, 295, 
80 



\bibitem[O'Dea 
\& Owen(1987)]{1987ApJ...316...95O} O'Dea, C.~P., \& Owen, F.~N.\ 1987, \apj, 316, 95 


\bibitem[O'Donoghue et al.(1993)]{1993ApJ...408..428O} O'Donoghue, A.~A., 
Eilek, J.~A., \& Owen, F.~N.\ 1993, \apj, 408, 428 

\bibitem[O'Sullivan et al.(2007)]{2007ApJ...658..299O} O'Sullivan, E., 
Vrtilek, J.~M., Harris, D.~E., \& Ponman, T.~J.\ 2007, \apj, 658, 299 

\bibitem[O'Sullivan et al.(2009)]{2009ApJ...701.1560O} O'Sullivan, E., 
Giacintucci, S., Vrtilek, J.~M., Raychaudhury, S., 
\& David, L.~P.\ 2009, \apj, 701, 1560 

\bibitem[O'Sullivan et al.(2010)]{2010MNRAS.407..321O} O'Sullivan, E., 
Giacintucci, S., David, L.~P., Vrtilek, J.~M., 
\& Raychaudhury, S.\ 2010, \mnras, 407, 321 


\bibitem[O'Sullivan et al.(2011)]{2011MNRAS.416.2916O} O'Sullivan, E., 
Worrall, D.~M., Birkinshaw, M., et al.\ 2011, \mnras, 416, 2916 

\bibitem[Pacholczyk(1970)]{1970ranp.book.....P} Pacholczyk, A.~G.\ 1970, 
Series of Books in Astronomy and Astrophysics, San Francisco: Freeman, 
1970


\bibitem[Parma et 
al.(2007)]{2007A&A...470..875P} Parma, P., Murgia, M., de Ruiter, H.~R., et al.\ 2007, \aap, 470, 875 

\bibitem[Quintana et 
al.(1994)]{1994A&A...283..722Q} Quintana, H., Fouque, P., \& Way, M.~J.\ 1994, \aap, 283, 722 


\bibitem[Randall et al.(2011)]{2011ApJ...726...86R} Randall, S.~W., Forman, 
W.~R., Giacintucci, S., et al.\ 2011, \apj, 726, 86 

\bibitem[Rasmussen 
\& Ponman(2007)]{2007MNRAS.380.1554R} Rasmussen, J., \& Ponman, T.~J.\ 2007, \mnras, 380, 1554 


\bibitem[Roettiger et al.(1996)]{1996ApJ...473..651R} Roettiger, K., Burns, 
J.~O., \& Loken, C.\ 1996, \apj, 473, 651 

\bibitem[Rudnick et al.(1994)]{1994ApJS...90..955R} Rudnick, L., 
Katz-Stone, D.~M., \& Anderson, M.~C.\ 1994, \apjs, 90, 955 




\bibitem[Saikia et al.(2006)]{2006MNRAS.366.1391S} Saikia, D.~J., Konar, 
C., \& Kulkarni, V.~K.\ 2006, \mnras, 366, 1391 



\bibitem[Saikia 
\& Jamrozy(2009)]{2009BASI...37...63S} Saikia, D.~J., \& Jamrozy, M.\ 2009, Bulletin of the Astronomical Society of India, 37, 63 



\bibitem[Schoenmakers et al.(2000)]{2000MNRAS.315..371S} Schoenmakers, 
A.~P., de Bruyn, A.~G., R{\"o}ttgering, H.~J.~A., van der Laan, H., 
\& Kaiser, C.~R.\ 2000, \mnras, 315, 371 

\bibitem[Subrahmanyan et al.(1996)]{1996MNRAS.279..257S} Subrahmanyan, R., 
Saripalli, L., \& Hunstead, R.~W.\ 1996, \mnras, 279, 257 



\bibitem[Tregillis et al.(2004)]{2004ApJ...601..778T} Tregillis, I.~L., 
Jones, T.~W., \& Ryu, D.\ 2004, \apj, 601, 778 


\bibitem[Trentham et al. (2006)]{2006MNRAS.369.1375T}
Trentham, N., Tully, R. B., Mahdavi, A. \ 2006, \mnras, 369, 1375

\bibitem[van Breugel 
\& Fomalont(1984)]{1984ApJ...282L..55V} van Breugel, W., \& Fomalont, E.~B.\ 1984, \apjl, 282, L55 

\bibitem[Weisskopf et al.(2002)]{2002PASP..114....1W} Weisskopf, M.~C., 
Brinkman, B., Canizares, C., et al.\ 2002, \pasp, 114, 1 


\bibitem[Wise et al.(2007)]{2007ApJ...659.1153W} Wise, M.~W., McNamara, 
B.~R., Nulsen, P.~E.~J., Houck, J.~C., 
\& David, L.~P.\ 2007, \apj, 659, 1153 

\end{thebibliography}
\end{document}